\documentclass[aps, nofootinbib]{revtex4}  

\usepackage[T1]{fontenc}
\usepackage[utf8]{inputenc} 
\usepackage{amsmath, amssymb, amsthm, bm, mathtools}
\usepackage{xcolor, dsfont}
\usepackage{hyperref}
\usepackage{datetime}
\usepackage{graphicx}
\usepackage{tikz}
\usetikzlibrary{decorations.pathmorphing, patterns}

\usepackage{algorithm}
\usepackage{algpseudocode}
\MakeRobust{\Call}

\newtheorem{theorem}{Theorem}
\newtheorem{lemma}{Lemma}
\newtheorem{corollary}{Corollary}
\theoremstyle{definition}
 
\newtheorem{definition}{Definition}
\newtheorem{remark}{Remark}

\newtheorem{construction}{Construction}

\DeclarePairedDelimiter\floor{\lfloor}{\rfloor}

\newcommand{\ket}[1]{| #1 \rangle}

\newcommand{\ketbra}[2]{|#1\rangle\!\langle#2|}

\definecolor{mylinkcolor}{rgb}{0,0,0.8}
\hypersetup{
  unicode=true,
  breaklinks=true,
  colorlinks=true,
  linkcolor=mylinkcolor,
  citecolor=mylinkcolor,
  urlcolor=mylinkcolor,
  filecolor=mylinkcolor
}

\usepackage[capitalise]{cleveref}
\usepackage{enumerate}
\usepackage[normalem]{ulem} 

\newcommand*{\Affaddr}[1]{\vspace{0.2cm}\centering\small\textit{#1}}
\newcommand*{\affaddr}[1]{\centering\small\textit{#1}}
\newcommand*{\affmark}[1][*]{\textsuperscript{#1}}

\begin{document}

\title{An efficient construction of Raz's two-source randomness extractor \\ with improved parameters}

\author{Cameron Foreman$^{\star}$\affmark[1,\,]\affmark[2,\,\textcolor{blue}{$*$}], Lewis Wooltorton$^{\star}$\affmark[3,\,]\affmark[4,\,]\affmark[5,\,\textcolor{blue}{$\dagger$}], Kevin Milner\affmark[1] and Florian J. Curchod\affmark[3] \\ 
\Affaddr{\affmark[1]Quantinuum, Partnership House, Carlisle Place, London SW1P 1BX, United Kingdom} \\
\affaddr{\affmark[2]Department of Computer Science, University College London, London, United Kingdom} \\
\affaddr{\affmark[3]Quantinuum, Terrington House, 13–15 Hills Road, Cambridge CB2 1NL, United Kingdom} \\
\affaddr{\affmark[4]Department of Mathematics, University of York, Heslington, York, YO10 5DD, United Kingdom} \\
\affaddr{\affmark[5]Quantum Engineering Centre for Doctoral Training, H. H. Wills Physics Laboratory and Department of Electrical \& Electronic Engineering, University of Bristol, Bristol BS8 1FD, United Kingdom}
}


\begin{abstract}
Randomness extractors are algorithms that distill weak random sources into near-perfect random numbers. Two-source extractors enable this distillation process by combining two independent weak random sources. 
Raz’s extractor (STOC '05) was the first to achieve this in a setting where one source has linear min-entropy (i.e., proportional to its length), while the other has only logarithmic min-entropy in its length.
However, Raz's original construction is impractical due to a polynomial computation time of at least degree 4. Our work solves this problem by presenting an improved version of Raz's extractor with quasi-linear computation time, as well as a new analytic theorem with reduced entropy requirements. We provide comprehensive analytical and numerical comparisons of our construction with others in the literature, and we derive strong and quantum-proof versions of our efficient Raz extractor. Additionally, we offer an easy-to-use, open-source code implementation of the extractor and a numerical parameter calculation module.
\end{abstract}

\maketitle

\def\thefootnote{$*$}\footnotetext{Electronic address: \href{mailto://cameron.foreman@quantinuum.com}{cameron.foreman@quantinuum.com}}
\def\thefootnote{$\dagger$}\footnotetext{Electronic address: \href{mailto://lewis.wooltorton@ens-lyon.fr}{lewis.wooltorton@ens-lyon.fr}}
\def\thefootnote{\arabic{footnote}}
\def\thefootnote{$\star$}\footnotetext{These authors contributed equally to this work.}\def\thefootnote{\arabic{footnote}}
\onecolumngrid

\section{Introduction}\label{sec:intro}

Perfectly unpredictable, or random, numbers are essential for applications such as numerical simulation and cryptography. However, generating such numbers directly is challenging, if not impossible. Most sources of random numbers produce outputs that are \textit{weakly random}, meaning that their outputs are only somewhat unpredictable. The most general weakly random source is a \textit{min-entropy source}~\cite{CG}, which, for sources that produce bitstrings of length $n$, has the guarantee that any output bitstring appears with a probability of at most $2^{-k}$, where $k$ is the source's \textit{min-entropy}. Min-entropy sources are common in practice because certifying significant structure on the unpredictability of the outcomes is typically difficult. This creates a problem: many applications require perfect randomness, yet most sources only produce weak randomness. Extensive research has focused on this issue, particularly through \textit{randomness extractors}, which distill weakly random sources into near-perfect random numbers.

Since min-entropy sources are the most general classification of random number generating sources, the ideal solution is to develop \textit{deterministic} randomness extractors that work for any min-entropy source. However, this is known to be impossible~\cite{CG} even when the source is almost perfect, i.e., has min-entropy deficient by just one bit ($k = n - 1$).
The next best approach is to construct \textit{probabilistic extractors}, which require an additional source of randomness.
Numerous probabilistic extractors exist, requiring different sources and assumptions. Broadly, they can be categorized into seeded extractors (requiring an additional \textit{seed} of perfect random numbers)~\cite{trevisan, Krawczyk, Hayashi16, Foreman_2024}, two-source extractors (requiring an additional min-entropy source)~\cite{Raz05, Bourgain_2005, Rao06,  Li_2012, Li_2016, Chattopadhyay_2016, Li_2023} and multi-source extractors (requiring multiple additional min-entropy sources)~\cite{goyal2021multi}. Of these, two-source extractors are the best solution, since they require the weakest assumptions and the least additional resources of probabilistic extractors. This makes them desirable for both theoretical and practical applications. Key examples include cryptography, where mismatches between theoretical assumptions and real-world conditions can lead to adversarial attacks, and in the derandomization of probabilistic algorithms,  where two-source extractors enable randomized algorithms to operate with an asymptotically vanishing amount of randomness \cite{K02}.

Recent theoretical advances in two-source extraction have resolved several long-standing open problems (see \cite{Chattopadhyay_2022} for a summary). However, a key outstanding challenge is whether these extractors can be implemented with a computation time suitable for real-world applications. Indeed, even quadratic-time $O(n^2)$ methods often become impractical for input sizes of $n \geq 10^6$, which are common in many practical scenarios. Appendix E of \cite{Hayashi16} gives a concrete example: using an $O(n^2)$ algorithm for privacy amplification in quantum key distribution (QKD) with an input size of $n=10^7$ can reduce throughput to at most 30 kbps, even on a 3 GHz clock-rate CPU processing 100 bits per cycle, which is far below the typical $\geq 300$ kbps in current QKD systems. We provide further numerical evidence of this observation in Section~\ref{subsec:performance}. Thus, it is essential to develop extraction algorithms with at most quasi-linear computation time, i.e., $O(n \log^k n)$ for some constant $k$.

Recently, the Dodis et al. two-source extractor \cite{Dodis_2004} was implemented with quasi-linear computation time \cite{Foreman_2023}, but it imposes strict constraints: both input sources are required to have equal length $n$, and the sum of their respective min-entropies, $k_1$ and $k_2$, must satisfy $k_1 + k_2 > n$. Raz's extractor~\cite{Raz05} relaxes these constraints, enabling extraction when one source (of length $n_1$) has linear min-entropy $k_1 > n_1/2$, and the other (of length $n_2$) has only logarithmic min-entropy $k_2 > O(\log n_2)$. However, the original algorithm runs in $O(n_1^4)$ time~\cite{Foreman_2023}, which poses a significant limitation. This raises the question of whether a more efficient, ideally quasi-linear time, implementation exists.

In this work, we solve this question by presenting an improved version of Raz's extractor, implemented with $O(n_1 \log(n_1)^2)$ computation time, with increased output length and reduced entropy requirements on the input sources. We provide both analytic and numerical parameter calculations across various security models, including extraction in the presence of an adversary with quantum side-information, and compare our results to other versions of Raz's extractor in the literature. In addition to theoretical improvements, we provide a highly optimized code implementation of the extractor, capable of handling input lengths up to $n_1 \approx 1.5 \cdot 10^8$, making it usable even in the device-independent regime~\cite{ColbeckThesis,CR_free,PABGMS,PAMBMMOHLMM}. Notably, we use the number-theoretic transform (NTT) instead of the fast Fourier transform (FFT) in our implementation, avoiding possible rounding errors caused by FFT floating point arithmetic\footnote{The FFT relies on floating-point arithmetic because it computes the discrete Fourier transform using complex roots of unity, which require floating-point approximations. This introduces rounding errors due to limited precision of representing and manipulating these numbers. In contrast, the NTT replaces complex roots with finite field roots of unity, allowing all operations to be performed exactly using modular arithmetic.}. Also provided is a separate calculation module that returns optimized extractor parameters based on a user-defined figure of merit, such as maximizing output length or minimizing entropy requirements. We demonstrate that this numerical approach outperforms known analytical theorems, due to the asymptotic statements of the theorems being non-optimal in finite-sized regimes. Both the parameter calculation module and the optimized implementation are publicly available in the Cryptomite library \cite{Foreman_2024} at \url{https://github.com/CQCL/cryptomite} and can be installed via the terminal command \texttt{pip install cryptomite}. The results presented in this work pave the way for future implementations of randomness amplification protocols (e.g., those which are device-independent~\cite{Kessler, Brandao, Foreman_2023, Ramanathan}), the efficient implementation of a broader class of extractors that use Raz's extractor as a subroutine (see, e.g., \cite{Kalai_09, Li_2012, Cohen, Cohen_16, Cohen_16b, Aggarwal, Aggarwal_2022, Goyal_2018, Goyal_2021, Chattopadhyay_2020, Li_2023}), and advancements in other tasks~\cite{Quach_2021, Dodis_2020, Ben_2020}.

The manuscript is structured as follows. In \cref{sec:background} we provide the necessary background. In \cref{sec:improving} we review the original statement of Raz's extractor and state our main results. \cref{sec:analysis} compares the performance of our efficient implementation and parameter calculation module, both with the original and other constructions in the literature. We then conclude and discuss some open problems in \cref{sec:discussion}. All proofs can be found in the Appendix.

\section{Background}\label{sec:background}
\subsection{Classical random variables}
We denote random variables using upper case, e.g., $X$, which take values $x$ in some finite alphabet $\mathcal{X}$ with probability $\mathrm{Pr}(X=x) = p_{X}(x)$. Given two random variables, $X$ and $Y$, over alphabets $\mathcal{X}$ and $\mathcal{Y}$ with distributions $p_{X}(x)$ and $p_{Y}(y)$, respectively, we label $X \circ Y$ the joint random variable over $\mathcal{X} \times \mathcal{Y}$ distributed as $p_{XY}(x, y)$ with marginals $p_{X}(x)$ and $p_{Y}(y)$. 
We label the distribution of $X$ conditioned on $Y$ by $p_{X|Y}(x|y) = \mathrm{Pr}(X=x|Y=y)$.
If $X$ and $Y$ are independent the joint distribution factors, i.e., $p_{XY}(x, y) = p_{X}(x)p_{Y}(y)$. Over a given finite alphabet (or domain) $\mathcal{D}$, the statistical distance between $X$ and $Y$ is $\mathrm{SD}[X, Y] = \frac{1}{2}\sum_{x \in \mathcal{D}}|p_{X}(x) - p_{Y}(x)|$. We denote by $p_{U_{n}}(x)$ the uniform distribution over an alphabet of size $2^{n}$ for some positive integer $n$, i.e., $p_{U_{n}}(x) = 2^{-n}$ when $\mathcal{X} = \{0, 1\}^{n}$. A random string of $n$ bits, $X = X_{0} \ldots X_{n-1}$, is then said to be $\epsilon$-close to uniform if $\mathrm{SD}[X, U_{n}] \leq \epsilon$.  Moreover, the min-entropy of $X$ is given by\footnote{Throughout this work, logarithms are taken to be base 2 unless otherwise stated.} $\mathrm{H}_{\infty}(X) = - \log [\max_{x \in \mathcal{X}} p_{X}(x)]$, and its min-entropy conditioned on $Y$ is $\mathrm{H}_{\infty}(X | Y) = - \log [\sum_{y \in \mathcal{Y}} p_Y(y) \max_{x \in \mathcal{X}} p_{X|Y}(x|y)]$. The random variable $X$ is called an $(n,k)$-source if it has a min-entropy $\mathrm{H}_{\infty}(X) \geq k$. For cryptographic applications, $X$ must have conditional min-entropy $\mathrm{H}_{\infty}(X|Y) \geq k$, where $Y$ is all side information accessible to the adversary.
The following definition is also needed:
\begin{definition}[$\zeta$-biased for linear tests of size $p'$]
    Let $Z = Z_{0}, \ldots, Z_{N-1}$ be an $N$-bit random variable. Let $p' \leq N$ be a positive integer, and $\zeta \geq 0$. 
    \begin{enumerate}[(a)]
        \item $Z$ is $\zeta$-biased for linear tests of size $p'$ if, for all non-empty subsets of indices $\tau \subseteq \{0, \ldots, N-1\}$ of size $|\tau| \leq p'$, the variable defined by $Z_{\tau} := \bigoplus_{i \in \tau} Z_{i}$ satisfies
        \begin{equation}
            2 \cdot \mathrm{SD}[Z_{\tau}, U_{1}] \leq \zeta\ .
        \end{equation}

        \item If $Z$ is $\zeta$-biased for linear tests of size $p' = N$, we say $Z$ is $\zeta$-biased for linear tests.

        \item Let $X$ be a bitstring of length $r < N$ distributed uniformly. A function $G:\{0,1\}^{r} \rightarrow \{0,1\}^{N}$ is a $(p',\zeta)$-biased generator if the random variable $G(X) = G(X)_{0}, \ldots, G(X)_{N-1}$ is $\zeta$-biased for linear tests of size $p'$.

        \item If $G$ is an $(N, \zeta)$-biased generator, we say $G$ is a $\zeta$-biased generator.
    \end{enumerate}
    \label{def:z-bias-p}
\end{definition}

\subsection{Two-source extractors}
We reproduce the following definition of a two-source extractor~\cite{Raz05}: 
\begin{definition}[Two-source extractor] \label{def:2-source}
    Let $X$ and $Y$ be any $(n_{1}, k_{1})$ and $(n_{2}, k_{2})$ independent sources, respectively. A function $\mathrm{Ext}:\{0,1\}^{n_{1}} \times \{0,1\}^{n_{2}} \rightarrow \{0,1\}^{m}$ that satisfies 
    \begin{equation}
        \mathrm{SD}[\mathrm{Ext}(X, Y), U_{m}] \leq \epsilon\ ,
    \end{equation}
    is called an $(n_{1}, k_{1}, n_{2}, k_{2}, m, \epsilon)$ two-source extractor. Moreover, $\mathrm{Ext}$ is said to be \textit{strong} in the first input if
    \begin{equation}
        \mathrm{SD}[\mathrm{Ext}(X, Y)\circ X, U_{m} \circ X] \leq \epsilon\ ,
    \end{equation}
    and strong in the second input if
    \begin{equation}
        \mathrm{SD}[\mathrm{Ext}(X, Y)\circ Y, U_{m} \circ Y] \leq \epsilon\ .
    \end{equation}
\end{definition}
\noindent Seeded extractors are a special case of two-source extractors in which $n_2=k_2$, meaning the second source is perfectly random and referred to as the \textit{seed}.

\subsection{Two-source extractors in the quantum setting with Markov sources}
\cref{def:2-source} can be generalized to sources that are independent in a weaker sense (e.g., under a Markov condition, see below) and can be made secure against adversaries capable of storing information in quantum systems. In the quantum setting, we denote the system $E$ as the adversary's (Eve's) quantum side-information, with the associated Hilbert space $\mathcal{H}_{E}$. 
Let $X$ and $Y$ be classical random variables which take values in $\{0,1\}^{n_1}$ and $\{0,1\}^{n_2}$, represented in Hilbert spaces $\mathcal{H}_{X}$ and $\mathcal{H}_{Y}$, respectively. The joint state with Eve before extraction is a classical-classical-quantum (ccq) state on $\mathcal{H}_{X} \otimes \mathcal{H}_{Y} \otimes \mathcal{H}_{E}$:
\begin{equation}
    \rho_{XYE} = \sum_{x \in \{0,1\}^{n_1}} \sum_{y \in \{0,1\}^{n_2}} p_{XY}(x, y) \, \ketbra{x}{x}_{X} \otimes \ketbra{y}{y}_{Y} \otimes \rho_{E}^{x,y}\ , \label{eq:ccq}
\end{equation}
where $\{\rho_{E}^{x,y}\}_{x,y}$ is a set of normalized quantum states on $\mathcal{H}_{E}$.
The minimum uncertainty in sampling $X$ or $Y$ from Eve's perspective is quantified by the conditional min-entropy $\mathrm{H}_{\infty}(X|E)_{\rho}$ or $\mathrm{H}_{\infty}(Y|E)_{\rho}$, respectively, evaluated on the state $\rho_{XYE}$.
For a classical-quantum (cq) state $\rho_{XE}$ on $\mathcal{H}_{X} \otimes \mathcal{H}_{E}$, the conditional min-entropy is given by $\mathrm{H}_{\infty}(X|E)_{\rho} = -\log[p_{\mathrm{guess}}(X|E)_{\rho}]$, where $p_{\mathrm{guess}}(X|E)_{\rho} = \max_{\{E_{x}\}_{x}} \sum_{x}\mathrm{Tr}[(\ketbra{x}{x}\otimes E_{x}) \rho_{XE}]$ and the maximum is taken over all positive operator-valued measures (POVMs) $\{E_{x}\}_{x}$ on $\mathcal{H}_{E}$.
In practice, the sources $X$ and $Y$ may be correlated. To account for this, we relax the independence assumption from the classical definition to that of conditionally independence given the adversary's information. In this case, $\rho_{XYE}$ is called a \textit{Markov source}:
\begin{definition}[Markov source]
    The ccq-state $\rho_{XYE}$ in \cref{eq:ccq} is a $(n_{1}, k_{1}), (n_{2}, k_{2})$ Markov source if $\mathrm{H}_{\infty}(X|E)_{\rho} \geq k_{1}$, $\mathrm{H}_{\infty}(Y|E)_{\rho} \geq k_{2}$ and $\mathrm{I}(X \, : \, Y | E)_{\rho} = 0$, where $\mathrm{I}(X : Y \mid E)_{\rho} = \mathrm{H}(XE)_{\rho} + \mathrm{H}(YE)_{\rho} - \mathrm{H}(XYE)_{\rho} - \mathrm{H}(E)_{\rho}$ denotes the conditional mutual information, and $\mathrm{H}(\cdot)_{\rho} = -\mathrm{Tr}[\rho \log \rho]$ is the von Neumann entropy.
    \label{def:markovS}
\end{definition}

Applying an extractor $\mathrm{Ext}: \{0,1\}^{n_1} \times \{0,1\}^{n_2} \rightarrow \{0,1\}^{m}$ to $XY$ can be described by a quantum channel $\mathcal{N}$ on $\rho_{XYE}$, where $\rho_{\mathrm{Ext}(X, Y)XYE} = \mathcal{N}[\rho_{XYE}]$. After tracing out $XY$, the state becomes
\begin{equation}
    \rho_{\mathrm{Ext}(X, Y)E} = \sum_{e \in \{0,1\}^{m}} p_{\mathrm{Ext}(X, Y)}(e) \, \ketbra{e}{e} \otimes \rho_{E}^{e}\ ,
\end{equation}
where $p_{\mathrm{Ext}(X, Y)}(e)\rho_{E}^{e} = \sum_{x, y | \mathrm{Ext}(x, y) = e} p_{XY}(x,y)\rho_{E}^{x, y}$. For two quantum states $\rho$ and $\sigma$ on a Hilbert space $\mathcal{H}$, we denote the trace distance by $\mathrm{TD}[\rho, \sigma] = \frac{1}{2}\|\rho - \sigma\|_{1} = \mathrm{Tr}[\sqrt{(\rho - \sigma)^{\dagger}(\rho - \sigma)}]$, which captures the maximum probability with which any quantum measurement can distinguish between $\rho$ and $\sigma$. We let $\omega_{m} = 2^{-m} \, \mathds{1}_{2^{m}}$ denote the maximally mixed state on $\mathcal{H} = \mathbb{C}^{2^{m}}$. We now define a two-source extractor in the quantum setting, considering Markov sources:
\begin{definition}[Quantum-proof two-source extractor secure in the Markov model]
     Let $n_{1},k_{1},n_{2},k_{2},m$ and $\rho_{XYE}$ be any $(n_{1},k_{1}),(n_{2},k_{2})$ Markov source. A function $\mathrm{Ext}:\{0,1\}^{n_{1}} \times \{0,1\}^{n_{2}} \rightarrow \{0,1\}^{m}$ which satisfies 
\begin{equation}
    \mathrm{TD}[\rho_{\mathrm{Ext}(X,Y)E},\omega_{m}\otimes \rho_{E}] \leq \epsilon\ ,
\end{equation}
where $\rho_{E} = \mathrm{Tr}_{\mathrm{Ext}(X,Y)}[\rho_{\mathrm{Ext}(X,Y)E}]$, is called a quantum-proof $(n_{1},k_{1},n_{2}, k_{2},m,\epsilon)$ two-source extractor secure in the Markov model.
Moreover, $\mathrm{Ext}$ is said to be strong in the first input $X$ if
\begin{equation}
    \mathrm{TD}[\rho_{\mathrm{Ext}(X,Y)XE},\omega_{m}\otimes \rho_{XE}] \leq \epsilon\ ,
\end{equation}
and strong in the second input $Y$ if 
\begin{equation}
    \mathrm{TD}[\rho_{\mathrm{Ext}(X,Y)YE},\omega_{m}\otimes \rho_{YE}] \leq \epsilon\ ,
\end{equation}
where $\rho_{XE}$ and $\rho_{YE}$ are defined by taking the partial trace of $\rho_{\mathrm{Ext}(X,Y)XYE}$. 
\end{definition}
\noindent Finally, we state the result of Arnon-Friedman et al.~\cite{AF15}, which shows that any two-source extractor can be made quantum-proof in the Markov model, albeit with a worse parameters:
\begin{lemma}[\cite{AF15}, Theorem 2]
    Every (strong) $(n_{1},k_{1},n_{2},k_{2},m,\epsilon)$ two-source extractor is a (strong) quantum-proof $(n_{1}, k_{1} + \log(1/\epsilon),n_{2},k_{2} + \log(1/\epsilon),m,\sqrt{3\epsilon2^{m-2}})$ two-source extractor secure in the Markov model. \label{lem:qproofMM}
\end{lemma}
\noindent Note that we place ``strong'' in parentheses because a strong two-source extractor retains this property when made quantum-proof in the Markov model, while a weak extractor (i.e., one that is not strong) remains weak.

\section{Improved Raz extractor}\label{sec:improving}

\subsection{The original construction}
In reference~\cite{Raz05}, Raz presents an explicit two-source extractor, which can be made strong in either input at the cost of a reduced output length.
Precisely, consider any independent $(n_{1},k_{1})$ source $X$ and $(n_{2},k_{2})$ source $Y$.
Let $m$ be a positive integer, $N = m \cdot 2^{n_{2}}$ and $G:\{0,1\}^{n_{1}} \rightarrow \{0,1\}^{N}$ be a $(p',\zeta)$-biased generator of output length $N$, as defined in \cref{def:z-bias-p}, i.e, if $X$  is distributed uniformly, the string $G(X)=G(X)_{0}, \ldots, G(X)_{N-1}$ is $\zeta$-biased for linear tests of size $p'$.
We can associate each generator output bit with a label $(i,y)$, where $i \in \{0,...,m-1\}$ and $y \in \{0,1\}^{n_{2}}$, and Raz shows that the function $\mathrm{Ext}:\{0,1\}^{n_{1}} \times \{0,1\}^{n_{2}} \rightarrow \{0,1\}^{m}$, defined bit-wise by $\mathrm{Ext}(x,y)_{i} = G(x)_{(i,y)}$, is a strong two-source extractor. 

\begin{lemma}[\cite{Raz05}, Lemma 3.3 and Lemma 3.4] Let $N = m \cdot 2^{n_{2}}$. Let $G_{0},...,G_{N-1}$ be 0-1 random variables that are $\zeta$-biased for linear tests of size $p'$, and can be constructed using $n_{1}$ random bits. Define $\mathrm{Ext}:\{0,1\}^{n_{1}} \times \{0,1\}^{n_{2}} \rightarrow \{0,1\}^{m}$ by $\mathrm{Ext}(x,y)_{i} = G(x)_{(i,y)}$. Then for any even integer $p \leq p'/m$, the function $\mathrm{Ext}$ is a $(n_1, k_1, n_2, k_2, m, \epsilon = 2^{m/2}\gamma)$ two-source extractor for any
\begin{equation}
    \gamma \geq 2^{(n_{1}-k_{1})/p} \cdot \big[ \zeta^{1/p} + p \cdot 2^{-k_{2}/2}\big]\ ,
\end{equation}
and a strong (in either input) $(n_{1},k_{1}',n_{2},k_{2}',m,\epsilon')$ two-source extractor with 
\begin{align}
    \nonumber k_{1}' &= k_{1} + m/2 + 2 + \log (1/\gamma)\ , \\
    k_{2}' &= k_{1} + m/2 + 2 + \log (1/\gamma)\ , \\
    \nonumber \epsilon' &= \gamma \cdot 2^{m/2+1}\ .
\end{align}
\label{lem:raz}
\end{lemma}

\noindent By an appropriate choice of $p$ and $ p'$ in \cref{lem:raz}, the following Lemma is recovered:
\begin{lemma}[\cite{Raz05}, Lemma 3.6]
    For any $n_{1},k_{1},n_{2},k_{2},m$ and any $0 < \delta' < 1/2$, such that
    \begin{equation}
        \begin{aligned}
            n_{1} &\geq 6 \log (n_{1}) + 2 \log (n_{2})\ , \\
            k_{1} &\geq (1/2 + \delta')n_{1} + 3\log (n_{1}) + \log(n_{2})\ , \\
            k_{2} &\geq 4 \log(n_{1}-k_{1})\ , \\
            m &\leq \delta'  \cdot \min[n_{1}/8,k_{2}/16] - 1\ ,
        \end{aligned}
    \end{equation}
    there exists an explicit strong $(n_{1},k_{1}',n_{2},k_{2}',m,\epsilon')$ two-source extractor with $\epsilon' = 2^{-3m/2}$, with 
    \begin{equation}
    \begin{aligned}
        k_1' &= k_1 + 3(m+1)\ , \\
        k_2' &= k_2 + 3(m+1)\ .
    \end{aligned}
    \end{equation}
\label{lem:raz2}
\end{lemma}

\noindent One can see that the above constraint on $k_1$ implies $k_{1}' \geq (1/2 + \delta')n_{1}$, i.e., the first source must have entropy rate $\alpha_{1} := k_{1}/n_{1}$ of at least $1/2$. For the second source, the third constraint implies that $k_{2}'$ can be logarithmic in $n_{1}$, implying that $\alpha_{2} := k_{2}/n_{2}$ can take values well below $1/2$. This allows Raz's extractor to break the barrier $\alpha_{1} + \alpha_{2} > 1$ required by the Dodis et al. extractor~\cite{Dodis_2004,Foreman_2023} and others \cite{Hayashi16, Berta_2021}. Furthermore, Raz's extractor requires an algorithm that generates variables biased for linear tests of size $p'$. The approach in \cite{Raz05} suggests using \cite[Lemma 4.1]{Naor93}, which comprises two algorithmic building blocks: $(i)$ generating strings $\zeta$-biased for linear tests \cite[Proposition 3]{Alon90}, and $(ii)$ generating strings which are $p'$-wise independent (that is, $(\zeta = 0)$-biased for linear tests of size $p'$) \cite[Proposition 6.5]{Alon86}. 
Based on these building blocks, it has been pointed out in \cite[Remark 14]{Foreman_2023} that, while the computation time for this implementation of Raz's extractor is polynomial in the input size $n_{1}$, it is at least $O(n_{1}^{4})$, making it unsuitable for most practical tasks.

\subsection{New construction with improved computation time}

To address the computation time bottleneck, we propose an implementation of Raz's extractor using the fast $(p',\zeta)$-biased generator due to Meka et al.~\cite{Meka14}. This construction does not rely on the concatenation of two steps, and its output can be computed in $O(\log(p'))$ finite field operations. 
Coupled with an algorithm for fast finite field arithmetic using circulant matrices~\cite{Hayashi16}, this approach reduces the overall computation time of Raz's extractor to $O(n_{1}\log(n_{1})\log(p'))$. In what follows, we will find that $p' = \mathrm{poly}(n_{1})$ is an appropriate choice for both our analytical and numerical parameter calculations (see \cref{thm:newRaz,sec:improving}), resulting in an overall quasi-linear computation time. 

\subsubsection{The fast $(p',\zeta)$-biased generator of reference~\cite{Meka14,comms}}
We present the construction from~\cite{Meka14} and its application as a computationally efficient $(p', \zeta)$-biased generator.
\begin{construction}[\cite{Meka14}, Section 1.1]
    Let $n $ and $ p'$ be positive integers satisfying $p' \leq n$. Let $\zeta > 0$ and $\mathbb{F}$ be a finite field with $|\mathbb{F}| \geq \mathrm{max}\{n,p'/\zeta\}$. Let $A,B$ be arbitrary subsets of $\mathbb{F}$, with $|A|=n$ and $|B| = p'/\zeta$. Define the generator $G:B \times \mathbb{F} \rightarrow \mathbb{F}^{|A|}$ as follows: for every $\alpha \in A$,
        \begin{equation}
            G(\beta,\nu)_{\alpha} := \nu \cdot \sum_{i=0}^{p'-1}(\alpha \beta)^{i}\ , \quad \beta \in B, \ \nu \in \mathbb{F}\ . \label{eq:gen}
        \end{equation}
        \label{const:eps}
\end{construction}

\noindent Let us fix the finite field $\mathbb{F}$ as the Galois field with $2^{t}$ elements, $\mathbb{F} = \mathrm{GF}[2^{t}]$, where $t$ is a positive integer. Suppose we choose $\zeta$ and $p'$ such that $r := \log(p'/\zeta)$ is a positive integer, and write $B = \mathrm{GF}[2^{r}]$. Then the input to the generator $G$ is an element $(\beta,\nu) \in \mathrm{GF}[2^{r}] \times \mathrm{GF}[2^{t}]$, which can be viewed as a bitstring of length $r+t$ generated from a uniform distribution. 
The output can be viewed as $|A|$ blocks of $t$ bits, or equivalently a bitstring of length $n \cdot t$ when $|A| = n$. Finally, according to Construction~\ref{const:eps}, we must choose values of $(r,t,n)$ such that $2^{t} \geq \max\{n,2^{r}\}$. 
We can therefore view $G$ as a $(p',\zeta)$-biased generator with a generator input (its \emph{seed}) of length $r+t$ bits and an output length of $n\cdot t$ bits. Moreover, reference~\cite{Meka14} continues to show that each block of $t$ bits (that is, the choice of $\alpha \in A$ in \cref{eq:gen}), can be computed efficiently. These facts are summarized below:  
\begin{lemma}[\cite{Meka14,comms}, Section 1.1]
    Let $n,\zeta$ and $p'$ be chosen according to Construction~\ref{const:eps} with $p'$ a positive power of $2$. Let $t$ be a positive integer, and suppose $r := \log(p'/\zeta)$ is a positive integer, such that $2^{t} \geq \max\{n,2^{r}\}$. Then the generator of Construction~\ref{const:eps} viewed as a function $G:\{0,1\}^{r+t} \rightarrow \{0,1\}^{n\cdot t}$ is a $(p',2\zeta)$-biased generator. Moreover, given any seed $(\beta,\nu) \in \mathrm{GF}[2^{r}] \times \mathrm{GF}[2^{t}]$ and an index $j \in \{0,...,n-1\}$, the $j^{th}$ block (of $t$ bits) can be computed using $O(\log (p'))$ field operations over $\mathrm{GF}[2^{t}]$.
    \label{lem:eff}
\end{lemma}
\noindent For completeness, we provide a detailed proof in \cref{app:genProof}. The efficiency claim comes from the following observation~\cite{Meka14}. Let $p' = 2^{l}$ where $l$ is a positive integer. Then
\begin{equation}
    G(\beta,\nu)_{\alpha} = \nu \cdot \sum_{i=0}^{p'-1}(\alpha \beta)^{i} = \nu \cdot \prod_{j=0}^{\log (p') - 1} (1 + (\alpha \beta)^{2^{j}})\ . \label{eq:eff}
\end{equation}
Since $\alpha,\beta,\nu \in \mathrm{GF}[2^{t}]$, one can verify the right hand side of \cref{eq:eff} can be computed in $O(\log (p'))$ finite field operations over $\mathrm{GF}[2^{t}]$. We emphasize that the generator $G$ is only efficient with respect to the computation of a single (or constant number) of blocks, rather than the entire output of the function.
\subsubsection{Application to Raz's two-source extractor}
We now match up the parameters of the generator to those required by Raz's extractor.
To summarize the above discussion, reference~\cite{Meka14} presents a $(p', 2\zeta)$-biased generator of output length $n \cdot t$, for some well chosen parameters $p',\zeta,n$ and $t$. The seed is of length $r + t$, where $r = \log(p'/\zeta)$, and output blocks of size $t$ can be computed in $O(\log (p'))$ field operations over $\mathrm{GF}[2^{t}]$. Firstly, the seed of $G$ should be the first source $X$, of length $n_{1}$, so we require $n_{1} = t + r$. Secondly, the number of output bits should be at least $m \cdot 2^{n_{2}}$, implying $n \cdot t \geq m \cdot 2^{n_{2}}$. A natural choice in Construction~\ref{const:eps} is to use the second source $Y$ to select the output block (that is, to choose $\alpha \in A$). The extractor output then corresponds to a subset of a single block, making it efficient to compute. This implies choosing $n = 2^{n_{2}}$, leaving us with the choice of $r$ and $t$ such that the constraints $(i)$ $n_{1} = t + r$,  $(ii)$ $t \geq m$, and $(iii)$ $2^{t} \geq \max\{2^{n_{2}},2^{r}\}$ are satisfied. Substituting $(i)$ into $(iii)$, we get $t \geq \max\{n_2, n_1 - t\}$, which implies $t \geq n_{1}/2$ 
and hence $r \leq n_{1}/2$. 
Since $r$ is proportional to $\log(1/\zeta)$, the best choice is to make $r$ as large as possible (to keep the bias $\zeta$ small), resulting in the symmetric construction $r = t = n_{1}/2$.
This finally implies $n_{1}/2 = \log(p'/\zeta)$, hence $\zeta = p' 2^{-n_{1}/2}$, leaving $p'$ as a free variable which is a power of $2$, i.e., $p' = 2^{l}$ for some positive integer $l$ to be specified later. Note that $p'$ must satisfy $p' \leq n \cdot t = (n_{1}/2)2^{n_{2}}$, hence $l \leq n_{2} + \log(n_{1}/2)$. Moreover, constraint $(ii)$ implies $n_{1}/2 \geq m$, and $(iii)$ further implies $n_{1}/2 \geq n_{2}$. We summarize this parameter matching as a Lemma:

\begin{lemma}[Efficient Raz's extractor construction] 
    Let $n_{1}$ and $n_{2}$ be positive integers, where $n_{1}$ is even and $n_{2} \leq n_{1}/2$. Define $N = (n_{1}/2)2^{n_{2}}$. Then for any positive integers $k_{1},k_{2},m,l,p$ and $\gamma > 0$ such that $m \leq n_{1}/2$, $l \leq n_{2} + \log(n_{1}/2), \ p \leq 2^{l}/m$, $p$ is even and any
    \begin{equation}
            \gamma \geq 2^{(n_{1}-k_{1})/p} \cdot \big[ (2\zeta)^{1/p} + p \cdot 2^{-k_{2}/2}\big],
        \end{equation}
    where $\zeta = 2^{l-n_{1}/2}$, we have the following:
    \begin{enumerate}[(i)]
        \item The generator of Construction~\ref{const:eps} viewed as a function $G:\{0,1\}^{n_{1}} \rightarrow \{0,1\}^{N}$ is a $(p',2\zeta)$-biased generator with $p' = 2^{l}$. 
        \item Consider the output of $G$ as $2^{n_{2}}$ blocks of size $n_{1}/2$ bits, and let $G(X)_{(i,y)}$ denote bit $i \in \{0,...,m-1\}$ of block $y \in \{0,1\}^{n_{2}}$. Then the function $\mathrm{Ext}:\{0,1\}^{n_{1}} \times \{0,1\}^{n_{2}} \rightarrow \{0,1\}^{m}$ defined by $\mathrm{Ext}(x,y)_{i} = G(x)_{(i,y)}$ is a $(n_1, k_1, n_2, k_2, m, \epsilon = 2^{m/2}\gamma)$ two-source extractor, and a strong (in either input) $(n_{1},k_{1}',n_{2},k_{2}',m, \gamma')$ two-source extractor, where
        \begin{equation}
            \begin{aligned}
                k_{1}' &= k_{1} + m/2 + 2 + \log (1/\gamma), \\
                k_{2}' &= k_{1} + m/2 + 2 + \log (1/\gamma), \\
                \gamma' &= \gamma \cdot 2^{m/2+1}.
            \end{aligned}
        \end{equation}
        \item Given $x \in \{0,1\}^{n_{1}}$ and $y \in \{0,1\}^{n_{2}}$, $\mathrm{Ext}(x,y)$ can be computed with computation time $O(n_{1}\log (n_{1})\log( p'))$.
    \end{enumerate}   \label{lem:newRaz1}
\end{lemma}
\noindent The proof of the above Lemma can be found in \cref{app:newRaz1}.
We now present a new version of \cite[Theorem 1]{Raz05}, giving an explicit Raz extractor with improved parameters that can be implemented in $O(n_{1}(\log n_{1})^{2})$ computation time. 
\begin{theorem}[Explicit and Efficient Raz Extractor]
    Let $n_{1}, k_{1}, n_{2}, k_{2}, m$ be positive integers, $0 < \delta < 1/2$ and $0.25 < \lambda < (\delta k_2 / 16 - 1)$, such that $n_{2} \leq n_{1}/2$ and
    \begin{align}\label{constr3_txt}
        k_1 &\geq \left(\frac{1}{2} + \delta\right)n_1 + 2\log(n_1)\ , \\
        \label{constr4_txt}
        k_2 &\geq \max\Big[3.2 \log \Big(\frac{8n_1}{k_2}\Big), 40 \Big]\ ,\\
       \label{constr5_txt}
        m &\leq \frac{1}{\lambda} \Big( \frac{\delta k_2}{16} - 1 \Big)\ .
    \end{align}
    Then there exists an explicit $(n_{1},k_{1},n_{2},k_{2},m,\epsilon \leq 2^{(1 - 4\lambda)m/2 - 1})$ two-source extractor, and an explicit strong (in either input) $(n_{1},k'_{1}, n_{2},k'_{2},m,\epsilon' \leq 2^{(1 - 4\lambda)m/2})$ two-source extractor, that can both be computed in $O(n_{1} \log(n_1)^2)$ time, with
    \begin{equation}
    \begin{aligned}
        k_1' &= k_1 + 3(m+1)\ , \\
        k_2' &= k_2 + 3(m+1)\ .
    \end{aligned}
    \end{equation} \label{thm:newRaz}
\end{theorem}
\noindent The proof of the above theorem can be found in \cref{app:newRaz}.

Selecting $\lambda = 1$ in \cref{thm:newRaz} (for the case $\delta k_{2}/16 - 1 > 1$) recovers an equivalent constraint on both the output length~\eqref{constr5_txt} and the error of Raz's original extractor from \cite[Lemma 3.6]{Raz05}, recalled here as \cref{lem:raz2}. For this choice of $\lambda$, our requirements on $k_1$ and $k_2$ are strictly weaker than those in \cref{lem:raz2} for any valid set of parameters $n_1, k_1, n_2, k_2$ and $m$ satisfying $n_2 \leq n_1/2$ and $k_2 \geq 8 / (1 - k_1/n_1)$. The first restriction arises because an efficient construction does not exist if $n_1 > n_2/2$ and the second restriction ensures that $3.2\log(8n_1/k_2) \leq 4 \log(n_1 - k_1)$, a condition that is almost always satisfied unless $k_1 \to n_1$.

Other works have also introduced improved analytic versions of Raz's extractor. In \cite{Foreman_2023}, the authors present an explicit, strong, and quantum-proof version of Raz's extractor. We compare this construction to those presented in this work in \cref{sec:analysis}. In~\cite[Section 5.2]{Aggarwal}, the authors propose a \textit{collision-resistant} variant of Raz's extractor, with strictly worse parameters than~\cite[Lemma 3.6]{Raz05}, and therefore generally performs worse than ours (as discussed above). Notably, the proof techniques from~\cite{Aggarwal} can be applied to our~\cref{thm:newRaz} to obtain an improved collision-resistant extractor, with better parameters and implementable in quasi-linear time.

\subsubsection{Making Raz's extractor quantum-proof in the Markov model}
Using \cref{thm:newRaz}, we now apply \cref{lem:qproofMM} to obtain a quantum-proof version of the efficient Raz extractor with improved parameters. 

\begin{corollary}[Efficient quantum-proof Raz extractor] 
     Let $n_{1}, n_{2}, k_{1}, k_{2}, m$ be positive integers, $0 < \delta < 1/2$ and $0.75 < \lambda < (\delta k_2 / 16 - 1)$, such that $n_{2} \leq n_{1}/2$ and
    \begin{align}
        k_1 &\geq \left(\frac{1}{2} + \delta\right)n_1 + 2\log(n_1) +1\ , \\
        k_2 &\geq \max\Big[3.2 \log \Big(\frac{8n_1}{k_2}\Big), 40 \Big]\ ,\\
        m &\leq \frac{1}{\lambda} \Big( \frac{\delta k_2}{16} - 1 \Big)\ .
    \end{align}
    Then there exists an explicit $(n_{1},k'_{1}, n_{2},k'_{2},m,\epsilon \leq \sqrt{3}\,2^{(3/4 - \lambda)m -1})$ strong (in either input) two-source extractor quantum-proof in the Markov model, which can be computed in $O(n_{1} \log( n_1)^2)$ time, where
    \begin{equation}
    \begin{aligned}
        k_1' &= k_1 + (2\lambda + 5/2)m + 3\ , \\
        k_2' &= k_1 + (2\lambda + 5/2)m + 3\ .
    \end{aligned}
    \end{equation} \label{cor:qproofRaz}
\end{corollary}
\noindent Note that we could also apply \cref{lem:qproofMM} directly to the efficient strong Raz extractor in \cref{lem:newRaz1}. The resulting parameters are less constrained than those in \cref{cor:qproofRaz}, as we retain flexibility in choosing $p$ and $p'$. These parameters can be optimized for a given problem, and this functionality is included in our parameter calculation module (see \cref{sec:analysis} for details). We summarize this construction below:

\begin{corollary}
    Let $N = m \cdot 2^{n_{2}}$. Let $G_{0},...,G_{N-1}$ be 0-1 random variables $\zeta$-biased for linear tests of size $p'$ that can be constructed using $n_{1}$ random bits. Define $\mathrm{Ext}:\{0,1\}^{n_{1}}\times \{0,1\}^{n_{2}} \to \{0,1\}^{m}$ by $\mathrm{Ext}(x,y)_{i} = G(x)_{(i,y)}$. Then, for any even integer $p \leq p'/m$ and any $k_{1},k_{2}$, the function $\mathrm{Ext}$ is a strong (in either input) quantum-proof $(n_{1},k_{1}',n_{2},k_{2}',m,2^{3m/4} \, \sqrt{3\gamma/2})$ two-source extractor in the Markov model, for any $\gamma \geq 2^{(n_{1}-k_{1})/p} \cdot \big[ \zeta^{1/p} + p \cdot 2^{-k_{2}/2}\big]$ and $k_{1}' = k_{1} + 1 + 2\log(1/\gamma)$, $k_{2}' = k_{2} + 1 + 2\log(1/\gamma)$. \label{cor:mbitQ2}
\end{corollary} 
\noindent 
While we have made the extractor quantum-proof in the Markov model, other models are also worth considering. For instance, in the model of~\cite{chung2014}, where side information is generated via a specific ``leaking operation'', Raz's extractor is known to be secure.

A crucial step in the proof of Raz's extractor is the application of the classical XOR lemma~\cite{GoldreichXOR}, which extends Raz's 1-bit extractor to an $m$-bit extractor. An alternative approach to making Raz's extractor quantum-proof would be to obtain a quantum-proof 1-bit extractor (e.g., using generic tools such as those in~\cite{Konig_2008,AF15}) and then apply a classical-quantum XOR lemma~\cite{Kasher}. This would generally yield a different set of parameters than those obtained by applying the generic approach (cf. \cref{lem:qproofMM}) to make the $m$-bit extractor quantum-proof in the Markov model. However, since the existing cq-XOR lemma~\cite{Kasher} is not as tight as its classical counterpart, we did not observe an improvement over the Markov model. Nevertheless, if a tighter cq-XOR lemma were proven, it could lead to better quantum-proof parameters. We refer the reader to \cref{app:cq1} for a more detailed discussion.

\subsubsection{Concatenation with a seeded extractor}
It is possible to increase the output length of Raz's extractor by using it in conjunction with a strong seeded extractor. Specifically, using Raz's strong extractor to generate $m_{\text{RAZ}}$ bits with error $\epsilon_{\text{RAZ}}$, and feeding this output into a seeded extractor, it is possible to re-extract from one of the original inputs. To enable re-extraction, it is important that the Raz extractor is strong (in the input used for re-extraction) and for $m_{\text{RAZ}}$ to be sufficiently long. If a strong seeded extractor is used, the output of the two-source extractor can be concatenated with the seeded extractor output, further increasing the output length. In the case of our improved construction, the logical choice is to re-extract from the first source (since $k_1 \geq k_2$ is always satisfied), which we summarize in the following remark.

\begin{remark} 
A strong Raz $(n_1, k_1, n_2, k_2, m_{\text{RAZ}}, \epsilon_{\text{RAZ}})$ two-source extractor and a strong $(n_1, k_1, d, d, m_{\text{S}}, \epsilon_{\text{S}})$ seeded extractor can be composed to obtain a $(n_1, k_1, n_2, k_2,  m_{\text{RAZ}} + m_{\text{S}}, \epsilon_{\text{RAZ}} + \epsilon_{\text{S}})$ two-source extractor if $m_{\text{RAZ}} \geq d$.
\end{remark}
\noindent For the proof, see e.g. \cite[Lemma 38]{AF15}.

A useful concatenation for our improved Raz's extractor is with the strong $(n_1, k_1, d, d, m_{\text{S}} = (c - 1) d, \epsilon_{\text{S}} = \sqrt{c-1} \cdot 2^{-d(1 + c(k_1/n_1 - 1))/2})$ seeded extractor, where $c$ is an integer such that $c \leq \lfloor \frac{1}{1 - k_1/n_1} \rfloor$, presented by Hayashi and Tsurumaru in~\cite{Hayashi16}. Notably, this extractor can be implemented in quasi-linear computation time (as detailed in \cite[App D.2]{Foreman_2023}) and has a non-vanishing output length when $c > 1$, i.e., when $k_1/n_1 > 0.5$, which is already a requirement of Raz's extractor. These facts, along with the fact that the output length scales as $\frac{1}{1 - k_1/n_1}$ rather than the seed length, makes the Hayashi-Tsurumaru extractor a good choice for the composition. For example, if $n_1/k_1 > 0.5$ we can always select $c=2$ and thus obtain an additional $d=m_{\mathrm{RAZ}}$ output bits (i.e., doubling the output length).

Another extractor to consider is Trevisan's \cite{trevisan}, which only requires a seed length of $O(\log(n_1))$ asymptotically (i.e., logarithmic in the length of the first source). An implementation of Trevisan's extractor was presented in \cite{Mauerer_2012, Foreman_2024}, giving a strong $(n_1, k_1, d, d, m_{\text{S}} = k_1 + 4\log(\epsilon_{\text{S}}) - 4\log(m_{\text{S}}) - 6, \epsilon_{\text{S}})$ seeded randomness extractor. Whilst the logarithmic seed length is a desirable property for composing with Raz's extractor, the drawback is that the length of $n_1$ needed to benefit from this asymptotic claim is typically substantial (except when the output length is very small). Moreover, the best-known implementations of Trevisan's extractor have a computation time of at least $O(n_1^2 \, \text{poly}(\log n_1))$, rendering them impractical for many applications. However, recent work by Doron and Ribeiro~\cite{doron2024nearly} proposes near-linear time constructions that, if implemented, could mitigate this bottleneck.

Both of the above compositions can be made quantum-proof by replacing only Raz's extractor with a quantum-proof version, as the extractors of Hayashi-Tsurumaru and Trevisan are quantum-proof without requiring any parameter changes~\cite{TSSR, DPVR}.

\subsection{Code implementation} \label{subsec:code}
We implement the Raz extractor in the Cryptomite library \cite{Foreman_2024}, following the technique described in \cite[Section~7.3.1]{DBLP:books/daglib/0021093}. This technique reduces finite field operations in $\mathrm{GF}[2^{n_1/2}]$ to polynomial convolutions in $\mathbf{Z}_w[x]/(x^{L}-1)$ for $w,L > n_1$, followed by the reduction by an irreducible polynomial of $\mathrm{GF}[2^{n_1/2}]$. These polynomial convolutions can be performed efficiently using the number-theoretic transform (NTT), and this transform is most efficient when the number of coefficients is a power of two; as such we perform convolutions using $w = 2^{32}$ and $L = {2^{\lceil\log(n_1)\rceil}}$ corresponding to the smallest suitable power of two and using 32-bit unsigned integer coefficients.

The main limitation of this technique is the need for an irreducible polynomial over the field $\mathrm{GF}[2^{n_1/2}]$, which can be time-consuming to find for large fields and is not generally known in advance. The Great Trinomial Hunt \cite{DBLP:journals/corr/abs-1005-1967} has identified irreducible trinomials for large fields $\mathrm{GF}[2^s]$ where $2^s-1$ is a Mersenne prime by exploiting the ability to efficiently test irreducibility when the factorization of $2^s-1$ is known, and exhaustively testing all possible trinomials for irreducibility. These are the largest fields for which irreducible polynomials are known, and so currently our technique is limited to $n_1/2 \leq 74,207,281$, i.e. $n_1 \approx 1.5 \cdot 10^8$, for which an irreducible trinomial is known.

The general procedure is shown in Algorithm~\ref{alg:raz}, where \textsc{NTT}() and \textsc{InvNTT}() implicitly pad the input with zeroes to the appropriate length, and $\odot$ denotes element-wise multiplication. The internal loop is well suited to parallelization, as $\zeta$ for iteration $j$ can be computed in parallel with $\delta_{cur}$ for iteration $j+1$. Furthermore, the two \textsc{NTT}() calls performed inside \textsc{Conv}() can be computed in parallel.

\begin{figure*}
\centering
\begin{minipage}{.7\linewidth}
\begin{algorithm}[H]
\caption{Raz Extractor.}\label{alg:raz}
\begin{algorithmic}
\Require $n_1\text{ where } \mathrm{GF}[2^{n_1/2}]\ \text{has a known irreducible polynomial $P$}$
\Require $p' \text{ where } p' \leq (n_1/2)2^{n_2}$ and $\log(p')$ is a positive integer
\Require $x \in \{0,1\}^{n_1}; y \in \{0,1\}^{n_2}; n_1 \geq 2 \cdot n_2; m \leq n_{1}/2$
\Statex
\Function{Extract}{$x$, $y$, $m$}
\State $x_1 \gets x[0:n_1/2]$
\State $x_2 \gets x[n_1/2:n_1]$
\State $\delta \gets \Call{Conv}{x_1,y}$ \Comment{$\delta_{0} = \alpha\beta^{2^0}$}
\State $\zeta \gets \delta + 1$ \Comment{$\zeta_0 = 1 + \alpha\beta^{2^0}$}
\For{$j \in [1, \dots, \log(p')-1]$}
    \State $\delta \gets \Call{Conv}{\delta_{cur}, \delta_{j}}$ \Comment{$\delta_{cur} = \alpha\beta^{2^j}$}
    \State $\zeta \gets \Call{Conv}{\zeta,\delta_{cur} + 1}$ \Comment{$\zeta_j = \zeta_{j-1} \cdot (1 + \alpha\beta^{2^j})$} 
\EndFor
\State \Return $\Call{Conv}{\zeta, x_2}[0:m]$
\EndFunction

\Statex
\Function{Conv}{$a$, $b$}
    \State $\Return\ \Call{InvNTT}{\Call{NTT}{a} \odot \Call{NTT}{b}} \mod P$
\EndFunction
\end{algorithmic}
\end{algorithm}
\end{minipage}
\end{figure*}

The code for our implementation and numerical parameter calculation module is available in the Cryptomite library (installable using terminal command \texttt{pip install cryptomite} or at \url{https://github.com/CQCL/cryptomite}).

\section{Analysis of the Improved Raz's extractor}\label{sec:analysis}
We now analyze the performance of our efficient Raz extractor and showcase it against alternative constructions.
Specifically, we analyze the maximal output length $m$ and the minimal possible entropy rate of the second source $\alpha_{2}$, given an extractor error $\epsilon$, a first source $(n_{1},k_{1})$ and a length $n_{2}$ of the second source. In our analysis, we do not compare the quantum-proof versions as all relevant works derive parameters using the same method (from \cite{AF15}), making such additional comparison redundant. For each optimization, we optimize over $p'$ and $p$ in \cref{lem:newRaz1} and all other parameters are fixed to correspond to regimes of interest. Some parameters are inherently constrained by our construction, such as $n_2 \leq n_1 / 2$.\footnote{We note that our calculations indicate that the optimal choice is $n_2 = n_1 / 2$.} Whether an efficient implementation of Raz's extractor exists for $n_2 > n_1 / 2$ remains an open question. Throughout the analysis, we also compare the performance when using our numerical optimization of parameters to the analytical version given in \cref{thm:newRaz} and the original Raz extractor in \cref{lem:raz2}. When relevant, we also compare with other implementations from the literature. 
\begin{remark}
    Whilst our approach using numerical optimization is tailored to the efficient extractor in this work, it could equally be applied to the original version in \cref{lem:raz}. \cref{fig:comp1,fig:comp2} show that our numerical parameter calculation leads to significantly better performance than the analytical theorems, and after a straightforward modification, one could also see the same benefits using \cref{lem:raz}. Moreover, the original construction has weaker constraints (for example, $p'$ is not restricted to be a power of two)
     and has a different expression for the error. The resulting performance will therefore at least match the one of our construction. Crucially though, the focus of this work is to provide a two-source extractor which can be implemented efficiently (in quasi-linear computation time), which is not achievable with the original Raz and other existing construction.      
\end{remark}
\noindent 

\subsection{Maximizing the output length}

\begin{figure}[h]
    \centering
    \includegraphics[width=0.43\textwidth]{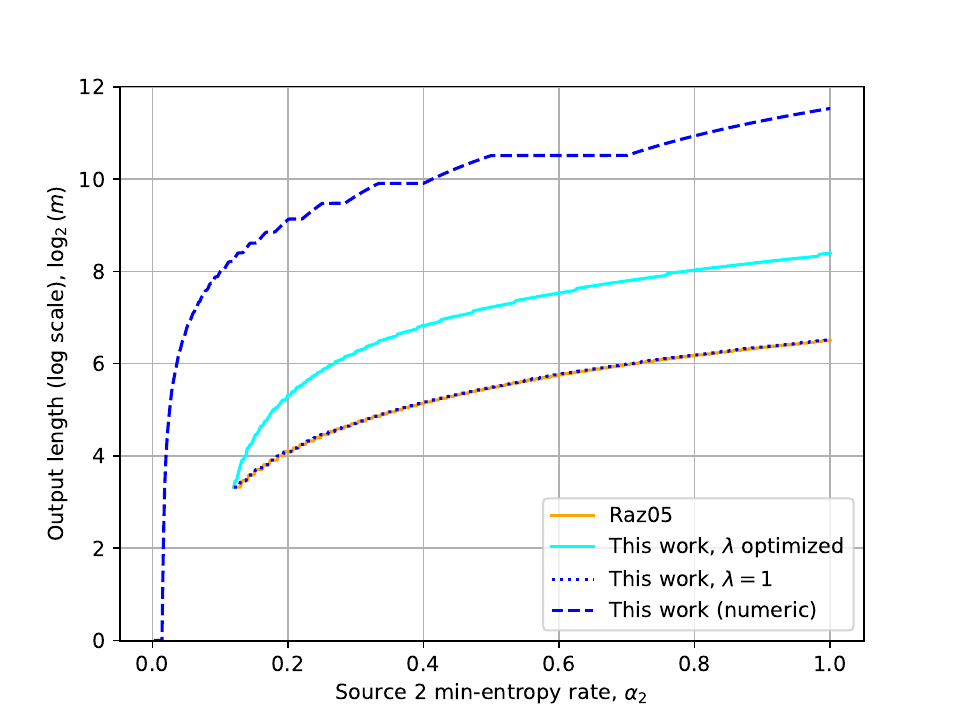}
    \includegraphics[width=0.43\textwidth]{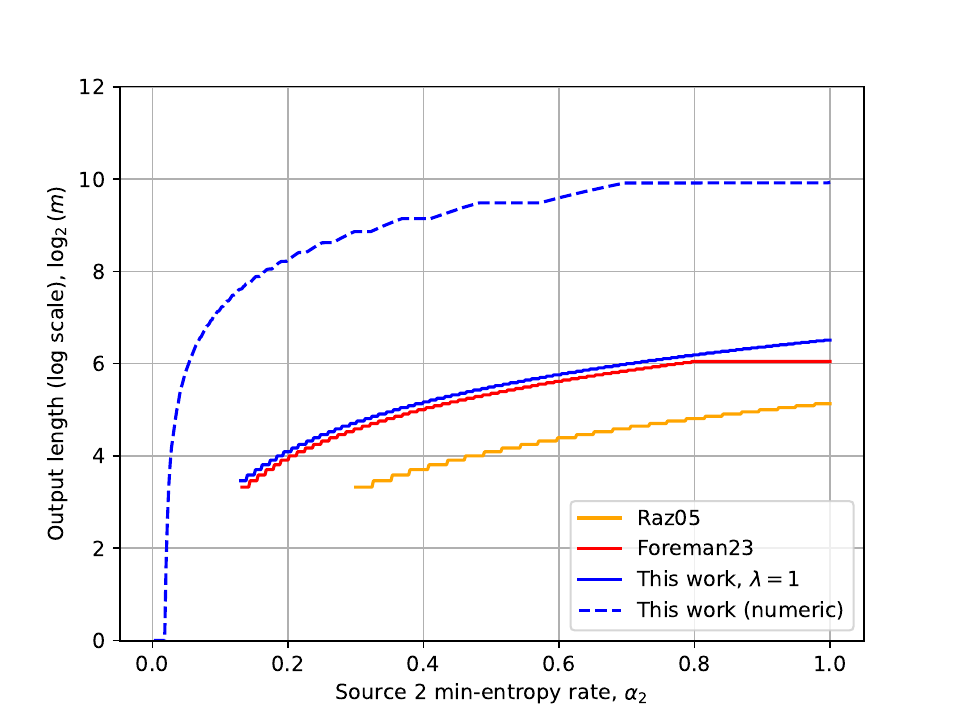}
    \caption{Comparison of maximum output lengths for different constructions of Raz's extractor, across different min-entropy rates of the second source, $\alpha_2$, with the extractor error $\epsilon \leq 2^{-16}$. We fix $n_{1} = 10^4$, $k_{1} = 0.8\times 10^{4}$ and $n_2 = n_1/2$. The left-hand side shows the weak extractor comparisons, while the right-hand side shows the strong comparisons. The comparison includes analytical (solid/dotted lines) and numerical optimization (dashed lines). Specifically, ``This work, $\lambda$ optimized'' and ``This work, $\lambda = 1$'' correspond to \cref{thm:newRaz}, while ``This work (numeric)'' corresponds to our numerical parameter calculation module using the default settings. The legend labels ``Raz05'' refers to the Raz extractor(s) from \cite{Raz05} and ``Foreman23'' refers to the strong Raz's extractor in \cite{Foreman_2023}.}
    \label{fig:comp1}
\end{figure}

To make comparisons, we fix the first source, $(n_{1},k_{1}) = (10^{4},0.8\times 10^{4})$, the length of the second source $n_{2} = n_{1}/2$, and set the extractor error to $\epsilon = 2^{-16}$. We then vary $\alpha_{2} \in (0,1]$ (recall that $k_{2} = \alpha_{2}n_{2}$) and maximize the output length $m$. In our numerical approach to parameter estimation, this optimization is performed over $p$ and $ p'$ satisfying the constraints in \cref{lem:newRaz1}. We consider both the weak and strong extractor constructions, and compare to our analytic values in \cref{thm:newRaz} with an optimized and fixed value of $\lambda$, as well as to the original Raz parameters in \cref{lem:raz2}. Our results are shown in \cref{fig:comp1}. 

We observe a significant improvement from our numerical analysis over the analytical theorems. This follows from the fact that the choice of $p$ and $ p'$ made analytically is often different from that found numerically, especially the choice of $p$. In the numerical calculations, we sometimes observe non-monotonic behavior of the maximum output length $m$ as a function of $\alpha_{2}$. When this occurs, we correct for it by considering the largest $m$ associated with any $\alpha_{2}' < \alpha_{2}$, since any $(n_{2}, \alpha_{2} n_{2})$ source is also a $(n_{2}, \alpha_{2}' n_{2})$ source.

We find that setting $\lambda = 1$ in our analytic theorem closely matches the performance of Raz's original theorem. We also see that, by optimizing over $\lambda$, our analytical theorem outperforms the original. This is because, by varying $\lambda$, we are able to keep the extractor error close to the fixed security parameter, rather than decreasing exponentially in $m$, as is the case for \cref{lem:raz2}. We observe that all analytical theorems require $m$ to be sufficiently large for any feasible choice of parameters (i.e., no feasible $\alpha_2$ yields a maximum output length smaller than three). This is because the extractor error in these versions is of the form $\approx 2^{cm}$, for some constant $c$, so a sufficiently long output length is necessary to achieve any desired error.

In the right-side plot of \cref{fig:comp1}, we compare our strong analytical version of Raz's extractor to the improved strong version presented in~\cite{Foreman_2023}. We find that our extractor performs similarly to the alternative in most regimes, but surpasses it as $\alpha_2$ approaches unity. This is because the output length in \cite{Foreman_2023} plateaus due to the requirement that $k_{2} < 2(n_{1} - k_{1})$, a constraint not present in our work. Therefore, our construction performs better in this regime. We note that this is a direct comparison, as we fix the errors to be equal (i.e., they are not a free parameter to be optimized over).

\subsection{Minimizing the entropy rate of the second source}
\begin{figure}[h]
    \centering
    \includegraphics[width=0.43\textwidth]{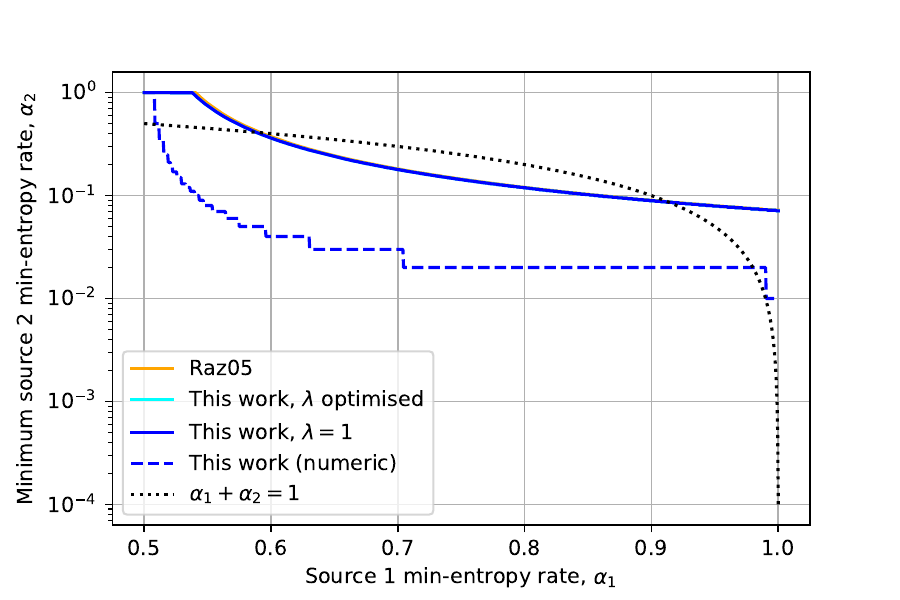}
    \includegraphics[width=0.43\textwidth]{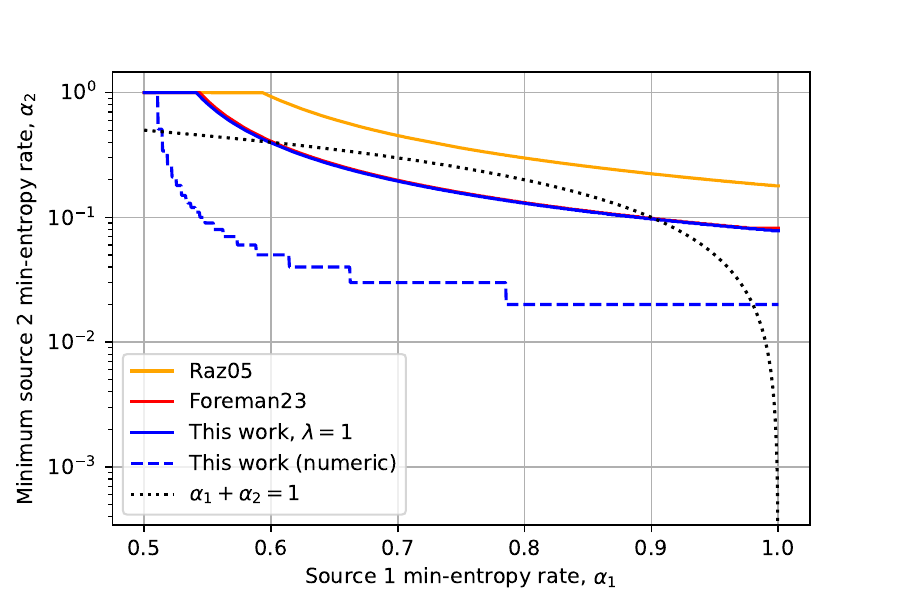}
    \caption{Comparison of the smallest min-entropy rate for the second source, $\alpha_2$, for which extraction is possible (i.e.\ $m \geq 1$) at different min-entropy rates of the first source, $\alpha_1$. We fix $n_{1} = 10^4$, $k_{1} = 0.8\times 10^{4}$ and $n_2 = n_1/2$. This is plotted for various constructions of Raz's extractor at an error tolerance of $\epsilon = 2^{-16}$. The left-hand side shows the weak extractor comparisons, while the right-hand side shows the strong comparisons. The comparison includes analytical (solid lines) and numeric (dashed lines). ``This work, $\lambda$ optimized'' and ``This work, $\lambda = 1$'' correspond to \cref{thm:newRaz}, while ``This work (numeric)'' corresponds to our numerical parameter calculation module using the default settings. The legend labels ``Raz05'' refers to the Raz's extractor(s) from \cite{Raz05} and ``Foreman23'' refers to the strong Raz's extractor in \cite{Foreman_2023}. The dotted line represents the theoretical limit of other efficient two-source extractors that are not based on Raz's construction, requiring $\alpha_1 + \alpha_2 > 1$.}
    \label{fig:comp2}
\end{figure}

We now consider the minimum entropy rate of the second source, $\alpha_{2}$, for which a single bit (i.e., $m = 1$) can be extracted. We fix the length of the first source $n_{1} = 10^{4}$, the extractor error to $\epsilon = 2^{-16}$ and vary $\alpha_{1} \in (0.5,1]$. Our numerical approach then minimizes $\alpha_{2}$ over feasible $p$ and $ p'$. The results are displayed in \cref{fig:comp2}. 

As with the maximization of the output length, our numerical approach yields a significant improvement over the analytical theorems. In our numerical approach, we also observe step-wise behavior in the minimum $\alpha_2$ as $\alpha_1$ increases. This occurs because $p$ and $p'$ must satisfy certain constraints, such as $p$ being even and $p'$ being a power of 2. We note that all versions can break the barrier $\alpha_1 + \alpha_2 > 1$ in the weak case, but in the strong case, Raz's original version fails to do so. This is due to the relatively small input lengths $n_1 = 10^4$, $n_2 = n_1/2$ and error requirement $\epsilon = 2^{-16}$.

In the weak case, we find that our analytic theorem closely matches the performance of Raz's original theorem, regardless of whether $\lambda$ is optimized. Optimizing $\lambda$ does not improve the outcome and simply recovers the same curve as when $\lambda = 1$. In the strong case, our theorem marginally outperforms that of \cite{Foreman_2023}, while both substantially improve upon Raz's original theorem. Again, we observe the plateau behavior of \cite{Foreman_2023} as $\alpha_2$ approaches unity, due to the additional constraint that $k_2 < 2(n_1 - k_1)$.

\subsection{Performance of the code implementation}
\label{subsec:performance}
\begin{figure}[H]
    \centering
    \includegraphics[width=0.35\textwidth]{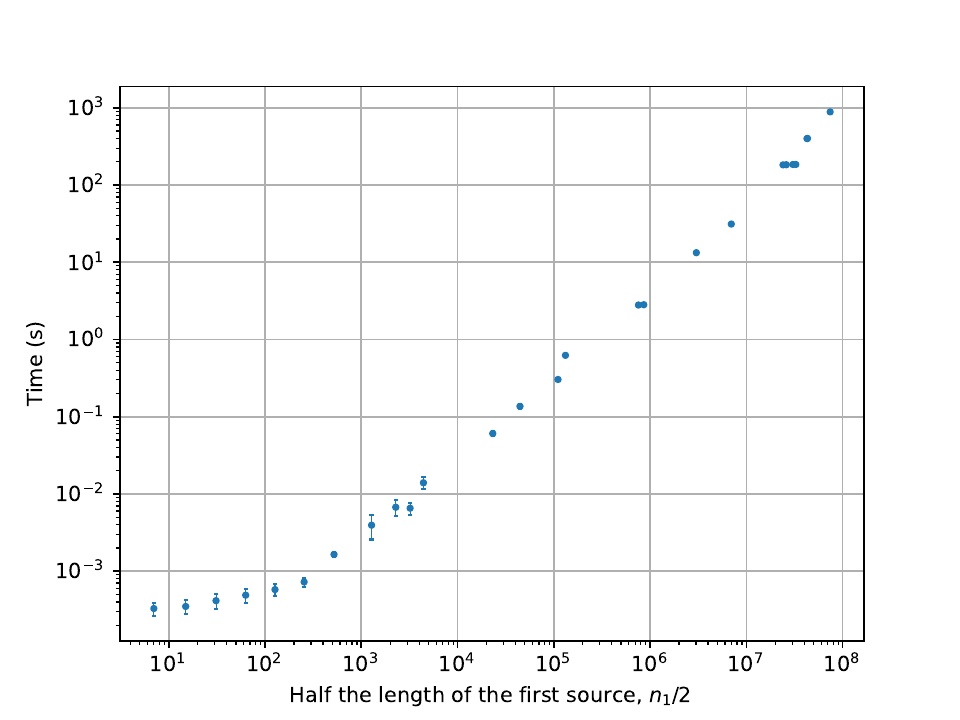}
    \caption{Benchmark results for the code implementation on an Apple M4 processor, showing the mean and two standard deviations over 20 runs.}
    \label{fig:benchmark}
\end{figure}

Recall from Section~\ref{subsec:code} that our technique requires a known irreducible polynomial for the field to compute field multiplications efficiently. We benchmarked our implementation using field sizes with known irreducible trinomials from the Great Trinomial Hunt~\cite{DBLP:journals/corr/abs-1005-1967}, up to the current maximum supported parameter $n_1/2 = 74,207,281$. We note that the runtime is independent of the output length and the choice of $n_2$. Our results are shown in Figure \ref{fig:benchmark}. These timings show that the expected quasi-linear runtime is achieved in practice, with small constant overhead dominating the runtime for small input lengths $n_1 \lessapprox 10^3$.

\subsection{Using Raz's extractor as a seeded extractor}
Another consideration is the performance of our construction as a seeded extractor, that is, when $k_i = n_i$ for $i \in \{1,2\}$. The only relevant scenario in our case arises when the seed is the second source, i.e., $k_2 = n_2$, as it is the shorter input. This constraint limits its usefulness as a seeded extractor, requiring the weak input to have a min-entropy rate of at least 0.5. However, we found that in some cases, the required seed length is shorter than that of other extractors with similar computation time, such as Hayashi-Tsurumaru \cite{Hayashi16}, Toeplitz \cite{Krawczyk}, and Circulant \cite{Foreman_2024}. Nonetheless, the substantial \textit{entropy loss} (the difference between the input min-entropy and output length) likely outweighs any potential advantage compared to these other seeded extractors.

\section{Discussion and Conclusion}\label{sec:discussion}
In this manuscript, we applied efficient techniques for constructing pseudo-random objects to Raz's two-source extractor~\cite{Raz05}. Specifically, Raz's extractor depends on generating bitstrings biased for linear tests, and existing implementations using~\cite{Alon90} suffer from a computational time of $O(n^{4})$, which is impractical. Using a more efficient algorithm from~\cite{Meka14}, we implemented Raz's extractor with a runtime of $O(n \log^2 n)$. As an additional contribution, we proved a new explicit theorem with entropy requirements lower than those of the original.

Our work opens a number of interesting research directions. The efficient implementation and accompanying code can be readily applied to the various use cases of Raz's extractor. In particular, Raz's extractor has weaker entropy requirements on one of its sources than other two-source extractors which have a known efficient implementation~\cite{Dodis_2004,Foreman_2023}. Randomness extractors are already known to be useful as exposure-resilient functions, for randomness extraction (or conditioning) of the output of noise sources, or to perform privacy amplification in quantum key distribution (QKD), where two-source extractors enable this under the weakest assumptions. This also implies the possibility of performing practical randomness amplification and privatization of weaker sources using quantum devices~\cite{CR_free}. In~\cite{tocome}, we perform randomness amplification of a single weak source that is a weakening of a Santha-Vazirani source \cite{SanthaVazirani} and our efficient Raz extractor construction significantly reduces the entropy requirements, allowing us to obtain new fundamental bounds. Another direction is to consider the efficient implementation of \textit{non-malleable extractors}, of which Raz's extractor is a common building block. Additionally, our efficient code implementation of the fast $(p',\zeta)$-biased generator from~\cite{Meka14} may be of independent interest for applications beyond randomness extraction.

In addition to applications, there are aspects of our construction that could be improved. For example, adapting the algorithm from~\cite{Meka14} introduced additional parameter constraints, such as the requirement $n_{2} \leq n_{1}/2$, and further restricted the choices of free parameters $p$ and $ p'$ when optimizing. It is an open question as to whether an efficient construction can be found that lifts these constraints, leading to greater versatility and performance. Finally, one could hope to reduce the penalties incurred by making Raz's extractor strong and quantum-proof. While we explored the use of certain classical-quantum XOR lemmas~\cite{GoldreichXOR,Kasher}, other techniques may yield improved parameters against quantum side information. For example, one could leverage the collision-resistance property from~\cite{Aggarwal}, apply the XOR lemma from~\cite{MSF}, or use the non-modular proof techniques developed there to show that the extractor by Dodis et al.~\cite{Dodis_2004} remains secure with the same parameters as in the classical setting.

\subsection*{Acknowledgements}
The authors thank Ron Rothblum for sharing the full version of~\cite{Meka14}, Sean Burton for reviewing the extractor code, and Kieran Wilkinson and Mafalda Almeida for valuable feedback on the manuscript. LW acknowledges support from the EPSRC Grant No. EP/SO23607/1.

\bibliographystyle{unsrt}
\bibliography{bibtex}

\onecolumngrid
\appendix

\section{Complete proofs for the fast $(p',\zeta)$-biased generator of reference~\cite{Meka14}} \label{app:genProof}
We now present the proof of \cref{lem:eff}. \textit{All results in this section of the appendix were obtained and proved in the full manuscript of~\cite{Meka14}. Following~\cite{Meka14,comms}, we reproduce the complete proofs for the convenience of the reader.} 

\vspace{0.2cm}

\noindent To begin, we introduce some extra notation and definitions. Let $\mathbb{F}$ be a finite field, and let $Z$ and $W$ be vectors in $\mathbb{F}^{n}$ with entries $Z_{i},W_{i} \in \mathbb{F}$ for $i \in \{0,...,n-1\}$, respectively. A vector $Z$ is $p'$-sparse if it has $p'$ non-zero entries. We denote the inner product over $\mathbb{F}^{n}$ by $\langle Z,W \rangle_{\mathbb{F}^{n}} = \sum_{i=0}^{n-1}Z_{i}  W_{i} \in \mathbb{F}$, where $Z_{i}  W_{i}$ denotes multiplication over $\mathbb{F}$. We denote $U_{\mathbb{F}^{n}}$ as a uniformly distributed random vector in $\mathbb{F}^{n}$, i.e., $p_{U_{\mathbb{F}^{n}}}(Z) = |\mathbb{F}|^{-n}$ for all $Z$ in $\mathbb{F}^{n}$. For the case $\mathbb{F} = \mathrm{GF}[2]$, we use the notation $\langle Z,W \rangle_{\text{bin}}$ for $\langle Z,W \rangle_{\mathrm{GF}[2]^{n}}$ and $U_{n}$ for $U_{\mathrm{GF}[2]^{n}}$, respectively (to maintain consistency with the main text).

\begin{definition}[$\zeta$-biased for linear tests of size $p'$ over $\mathbb{F}$]
    Let $Z$ be a random vector in $\mathbb{F}^{n}$. Let $p' \leq n$ be a positive integer and $\zeta \geq 0$. $Z$ is $\zeta$-biased for linear tests of size $p'$ over $\mathbb{F}$ if, for all non-zero $p'$-sparse vectors $W \in \mathbb{F}^{n}$, the variable defined by $Z_{W} := \langle Z,W \rangle_{\mathbb{F}^{n}}$ satisfies
        \begin{equation}
            2 \cdot \mathrm{SD}[Z_{W},U_{\mathbb{F}}] \leq \zeta \ .
        \end{equation} \label{def:biasfield}
\end{definition}
\noindent Let $X$ be a bitstring of length $r$ distributed uniformly. A function $G:\{0,1\}^{r} \to \mathbb{F}^{n}$ is a $(p',\zeta)$-biased generator over $\mathbb{F}$ if the variable $G(X)$ is $\zeta$-biased for linear tests of size $p'$ over $\mathbb{F}$. We also restate the construction below for convenience.

\vspace{0.2cm}

\noindent \textbf{Construction 1} (\cite{Meka14,comms}, Section 1.1).
    Let $n $ and $ p'$ be positive integers that satisfy $p' \leq n$. Let $\zeta > 0$ and $\mathbb{F}$ be a finite field with $|\mathbb{F}| \geq \mathrm{max}\{n,p'/\zeta\}$. Let $A,B$ be arbitrary subsets of $\mathbb{F}$, with $|A|=n$ and $|B| = p'/\zeta$. Define the generator $G:B \times \mathbb{F} \rightarrow \mathbb{F}^{|A|}$ as follows: for every $\alpha \in A$,
        \begin{equation}
            G(\beta,\nu)_{\alpha} := \nu \cdot \sum_{i=0}^{p'-1}(\alpha \beta)^{i}\ , \quad \beta \in B, \ \nu \in \mathbb{F} \ . 
        \end{equation}

\begin{lemma}[\cite{Meka14,comms}, Section 1.1]
    $G$ as defined in Construction~\ref{const:eps} is a $(p',2\zeta)$-biased generator over $\mathbb{F}$ according to \cref{def:biasfield}.
    \label{lem:fieldBias}
\end{lemma}
\begin{proof}
    The generator in Construction~\ref{const:eps} maps finite field elements in $B \times \mathbb{F}$ to a vector in $\mathbb{F}^{|A|}$. Let $W$ be any vector in $\mathbb{F}^{|A|}$ with entries indexed by $W_{\alpha}$ for $\alpha \in A$. Then, according to \cref{def:biasfield}, we want to show that the random variable 
    \begin{equation}
        G_{W} := \langle W , G(\beta,\nu) \rangle_{\mathbb{F}} \ ,
    \end{equation}
    is statistically close to $U_{\mathbb{F}}$. Let $\lambda \in \mathbb{F}$, and define the degree $p'-1$ polynomial over $\mathbb{F}$,
    \begin{equation}
        P_{W}(\lambda) := \sum_{i=0}^{p'-1}\Bigg( \sum_{\alpha \in A} W_{\alpha} \alpha^{i}\Bigg) \lambda^{i} \ .
    \end{equation}
    Then, for every pair of generator inputs $\beta \in B$ and $\nu \in \mathbb{F}$,
    \begin{equation}
        \langle W , G(\beta,\nu) \rangle_{\mathbb{F}} = \sum_{\alpha \in A} W_{\alpha} G(\beta,\nu)_{\alpha} = \sum_{\alpha \in A} \sum_{i = 0}^{p'-1} W_{\alpha} \cdot \nu \cdot (\alpha \beta)^{i} = \nu \cdot \sum_{i=0}^{p'-1} \Bigg( \sum_{\alpha \in A} W_{\alpha} \alpha^{i} \Bigg) \beta^{i} = \nu \cdot P_{W}(\beta) \ .
    \end{equation}

    Next, we notice that $P_{W}$ is a non-zero polynomial whenever $W$ is $p'$-sparse. To see this, let $\tau = \{t_{0},...,t_{p'-1}\} \subseteq A$ denote the set of coordinates $\alpha$ for which $W_{\alpha}$ is non-zero. The coefficients $c_{i} \in \mathbb{F}$ of $P_{W}(\lambda)$ can be expressed as the inner product
    \begin{equation}
        c_{i} := \sum_{\alpha \in A} W_{\alpha} \alpha^{i} = \sum_{\alpha \in \tau} W_{\alpha} \alpha^{i} = \Big\langle [W_{t_{0}},...,W_{t_{p'-1}}] , [(t_{0})^{i},...,(t_{p'-1})^{i}] \Big \rangle_{\mathbb{F}} \ ,
    \end{equation}
    for $i = 0,...,p'-1$. The vector $[c_{0},...,c_{p'-1}]$ can therefore be expressed as the vector-matrix product
    \begin{equation}
        [c_{0},...,c_{p'-1}] = [W_{t_{0}},...,W_{t_{p'-1}}] \times 
        \underbrace{\begin{bmatrix}
            1 & t_{0} & (t_{0})^{2} & \cdots & (t_{0})^{p'-1}\\
            1 &  t_{1} & (t_{1})^{2} & \cdots & (t_{1})^{p'-1} \\
            \vdots & \vdots & \vdots & \ddots & \vdots \\
            1 & t_{p'-1} & (t_{p'-1})^{2} & \cdots & (t_{p'-1})^{p'-1}
        \end{bmatrix}}_{\textbf{V}}\ ,
    \end{equation}
    The matrix $\textbf{V} \in \mathbb{F}^{\, p' \times p'}$ above is a Vandermonde matrix, which is invertible if and only if all $\{t_{i}\}_{i=0}^{p'-1}$ are distinct. This condition is satisfied by definition, as they correspond to the indices of $W$ with non-zero entries. Therefore, $[c_{0}, \ldots, c_{p'-1}] \mathbf{V}^{-1} = [W_{t_{0}}, \ldots, W_{t_{p'-1}}]$ and, since $[W_{t_{0}}, \ldots, W_{t_{p'-1}}]$ is non-zero by definition, the vector $[c_{0}, \ldots, c_{p'-1}]$ cannot be the all zero vector.

    Consequently, since $P_{W}$ is a non-zero polynomial of degree at most $p'-1$ over $\mathbb{F}$, it has at most $p'-1$ roots. Given that $\beta$ is chosen uniformly from $B$, with $|B| = p'/\zeta$, the probability that $P_{W}(\beta) = 0$ is at most $(p'-1)/|B| = \zeta - \zeta/p' \leq \zeta$. Conditioned on $P_{W}(\beta) \neq 0$, the variable $\nu \cdot P_{W}(\beta)$ is uniformly distributed over $\mathbb{F}$ (since $\nu$ is chosen uniformly over $\mathbb{F}$). We therefore have
    \begin{equation}
    \begin{aligned}
         \mathrm{SD}[G_{W},U_{\mathbb{F}}] &= \frac{1}{2}\sum_{\mu \in \mathbb{F}} \Big | p_{G_{W}}(\mu) - \frac{1}{|\mathbb{F}|} \Big| 
         = \frac{1}{2}\sum_{\mu \in \mathbb{F}} \Big | \mathrm{Pr}[\nu \cdot P_{W}(\beta) = \mu ] - \frac{1}{|\mathbb{F}|} \Big|\\
         &= \frac{1}{2}\sum_{\mu \in \mathbb{F}} \Big | \mathrm{Pr}[P_{W}(\beta) = 0]\mathrm{Pr}[\nu \cdot P_{W}(\beta) = \mu| P_{W}(\beta) = 0] \\ 
         & \hspace{1.2cm} + (1-\mathrm{Pr}[P_{W}(\beta) = 0])\mathrm{Pr}[\nu \cdot P_{W}(\beta) = \mu| P_{W}(\beta) \neq 0] 
          - \frac{1}{|\mathbb{F}|} \Big|\\
          &= \frac{1}{2}\sum_{\mu \in \mathbb{F}} \Big | \mathrm{Pr}[P_{W}(\beta) = 0]\delta_{0,\mu}  + \frac{(1-\mathrm{Pr}[P_{W}(\beta) = 0])}{|\mathbb{F}|} - \frac{1}{|\mathbb{F}|} \Big|\\ 
          &= \frac{1}{2}\sum_{\mu \in \mathbb{F}} \Big | \mathrm{Pr}[P_{W}(\beta) = 0]\Big( \delta_{0,\mu} - \frac{1}{|\mathbb{F}|} \Big)  \Big| \\
          &\leq \frac{\zeta}{2} \Big | 1 - \frac{1}{|\mathbb{F}|} \Big | + \frac{\zeta}{2} \sum_{\mu \in \mathbb{F} \, : \, \mu \neq 0}   \frac{1}{|\mathbb{F}|}  = \zeta(1-1/|\mathbb{F}|) \leq \zeta \ .
    \end{aligned}
    \end{equation}
    For the fourth equality, we used the facts $\mathrm{Pr}[\nu \cdot P_{W}(\beta) = \mu| P_{W}(\beta) \neq 0]  = 1/|\mathbb{F}|$ and $\mathrm{Pr}[\nu \cdot P_{W}(\beta) = \mu| P_{W}(\beta) = 0] = \delta_{0,\mu}$. For the first inequality, we used $\mathrm{Pr}[P_{W}(\beta) = 0] \leq \zeta$ and separated the $\mu = 0$ from the sum. Multiplying both sides by 2 completes the proof.
\end{proof} 

Given a bitstring $X \in \{0,1\}^{m}$, we define the function $\mathrm{LSB}:\{0,1\}^{m} \to \{0,1\}$, which returns the least significant bit of $X$, i.e. if $X = X_{0},...,X_{m-1}$, $\mathrm{LSB}(X) = X_{m-1}$. We identify elements of $\mathrm{GF}[2^{m}]$ with vectors in $\mathrm{GF}[2]^{m}$ in the natural way\footnote{By this, we mean an element $Z \in \mathrm{GF}[2]^{m}$ is viewed as a bitstring in $\{0,1\}^{m}$, where the $i$-th bit is identified with coefficient $i-1$ of the polynomial representation of $Z$.} (and vice-versa) and, for two bitstrings $X$ and $Y$ of length $m$, $X + Y$ denotes their addition over $\mathrm{GF}[2^{m}]$ (represented as a bitstring). Then, we have that $\mathrm{LSB}(X+Y) = \mathrm{LSB}(X) \oplus \mathrm{LSB}(Y)$. Moreover, for every $\alpha \in \mathrm{GF}[2^{m}]$, define $T_{\alpha} : \mathrm{GF}[2]^{m} \to \mathrm{GF}[2]$ by $T_{\alpha}(\lambda) = \mathrm{LSB}(\alpha \cdot \lambda)$. Note that $T_{\alpha}$ is linear, and we view the product $\alpha \cdot \lambda$ as multiplication over $\mathrm{GF}[2^{m}]$. We have the following:

\begin{lemma}[\cite{Lidl96}, Theorem 2.24]
    For every $\alpha \in \mathrm{GF}[2^{m}]$, there exists an $\alpha' \in \mathrm{GF}[2^{m}]$ such that $T_{\alpha'}(\lambda) = \langle \alpha, \lambda \rangle_{\text{bin}}$ for all $\lambda \in \mathrm{GF}[2]^{m}$. \label{lem:tmap}
\end{lemma}

\begin{lemma}[\cite{Meka14,comms}, Section 1.1]
    Let $G: \{0,1\}^{r} \to \mathrm{GF}[2^{m}]^{n}$ be a $(p',\zeta)$-biased generator over $\mathrm{GF}[2^{m}]$. Then $G$ viewed as a function from $\{0,1\}^{r} \to \mathrm{GF}[2]^{m \cdot n}$ is a $(p',\zeta)$-biased generator over $\mathrm{GF}[2]$.  \label{lem:con}
\end{lemma}
\begin{proof}
    In the following, given a vector $V \in \mathrm{GF}[2]^{m\cdot n}$, we denote its entries by $V_{j}$ for $j \in \{0,...,nm-1\}$. We can also view $V$ as $n$ blocks of size $m$, and we write $\bar{V}_{i}$, $i \in \{0,...,n-1\}$ for each block, with $\bar{V}_{i} \in \mathrm{GF}[2]^{m}$. In other words, we use bars to denote blocks of size $m$, and $\bar{V}_{i}$ is the $i$-th block of $V$. Let $W \in \mathrm{GF}[2]^{m\cdot n}$ be any non-zero $p'$-sparse vector, and $X$ be a uniformly distributed bitstring over $\{0,1\}^{r}$. We will now show that the variable $G_{W} := \langle W, G(X) \rangle_{\text{bin}} \in \mathrm{GF}[2]$ is close to $U_{1}$ in statistical distance. 

    For every $W$, consider the block $\bar{W}_{i} \in \mathrm{GF}[2]^{m}$. Let $\bar{W}_{i}'\in \mathrm{GF}[2]^{m}$ be defined such that $T_{\bar{W}_{i}'}(\lambda) = \langle \bar{W}_{i} , \lambda \rangle_{\text{bin}}$ for all $\lambda \in \mathrm{GF}[2]^{m}$, according to \cref{lem:tmap}. Now, we construct the vector $W' \in \mathrm{GF}[2^{m}]^{n}$ with entries $\bar{W}_{i}'$. If $W$ is non-zero and $p'$-sparse over $\mathrm{GF}[2]$, then $W'$ is also non-zero and $p'$-sparse over $\mathrm{GF}[2]$. Given any vector $S \in \mathrm{GF}[2]^{m \cdot n}$, we have
    \begin{equation}
        \langle W,S \rangle_{\text{bin}} = \bigoplus_{i=0}^{n-1}\langle \bar{W}_{i}, \bar{S}_{i} \rangle_{\text{bin}} = \bigoplus_{i=0}^{n-1} \mathrm{LSB}\Big(\bar{W}_{i}' \cdot \bar{S}_{i} \Big) =  \mathrm{LSB}\Bigg(\sum_{i=0}^{n-1} \bar{W}_{i}' \cdot \bar{S}_{i} \Bigg) = \mathrm{LSB}\Big(\big \langle W' , S \big \rangle_{\mathrm{GF}[2^{m}]} \Big) \ .
    \end{equation}
    Applying the above to $S = G(X)$, we have
    \begin{equation}
        \langle W,G(X) \rangle_{\text{bin}} = \mathrm{LSB}\Big(\big \langle W' , G(X) \big \rangle_{\mathrm{GF}[2^{m}]} \Big) \ .
    \end{equation}
    Since $G$ is a $(p',\zeta)$-biased generator over $\mathrm{GF}[2^{m}]$, the variable $\langle W' , S \big \rangle_{\mathrm{GF}[2^{m}]}$ for any non-zero $p'$-sparse vector $W'$ is distributed $\zeta$-close to uniform over $\mathrm{GF}[2^{m}]$. Therefore, when viewing $\langle W' , S \big \rangle_{\mathrm{GF}[2^{m}]} \in \mathrm{GF}[2]^{m}$ as an $m$ length bitstring, every individual bit is distributed $\zeta$-close to uniform over $\mathrm{GF}[2]$. Therefore, $\langle W,G(X) \rangle_{\text{bin}}$ is distributed $\zeta$-close to uniformly over $\mathrm{GF}[2]$, proving the claim. 
\end{proof}

\noindent We can now establish \cref{lem:eff}.

\vspace{0.2cm}

\noindent \textbf{Lemma 4} (\cite{Meka14,comms}, Section 1.1).
\textit{Let $n,\zeta$ and $p'$ be chosen according to Construction~\ref{const:eps} with $p'$ a positive power of $2$. Let $t$ be a positive integer, and suppose $r := \log(p'/\zeta)$ is a positive integer, such that $2^{t} \geq \max\{n,2^{r}\}$. Then the generator of Construction~\ref{const:eps} viewed as a function $G:\{0,1\}^{r+t} \rightarrow \{0,1\}^{n\cdot t}$ is a $(p',2\zeta)$-biased generator. Moreover, given any seed $(\beta,\nu) \in \mathrm{GF}[2^{r}] \times \mathrm{GF}[2^{t}]$ and an index $j \in \{0,...,n-1\}$, the $j^{th}$ block (of $t$ bits) can be computed using $O(\log (p'))$ field operations over $\mathrm{GF}[2^{t}]$.}

\begin{proof}
    Based on \cref{lem:fieldBias}, we know $G: B \times \mathbb{F} \to \mathbb{F}^{|A|}$ is a $(p',2\zeta)$-biased generator over $\mathbb{F}$, for well chosen $\mathbb{F},A,B,p',n$ and $\zeta$ according to Construction~\ref{const:eps}. Let us choose $\mathbb{F} = \mathrm{GF}[2^{t}]$, $B = \mathrm{GF}[2^{r}]$ with $r = \log(p'/\zeta)$, and $|A| = n$. Then $G:\mathrm{GF}[2^{r}] \times \mathrm{GF}[2^{t}] \to \mathrm{GF}[2^{t}]^{n}$ is a $(p',2\zeta)$-biased generator over $\mathrm{GF}[2^{t}]$ when the condition $|\mathbb{F}| \geq \max\{|A|,|B|\}$ is satisfied, which translates to $2^{t} \geq \max\{n,2^{r}\}$. Now, viewing $G$ as a function $G:\{0,1\}^{r+t} \to \mathrm{GF}[2]^{n\cdot t} \equiv \{0,1\}^{n \cdot t}$, we can apply \cref{lem:con} to show $G$ is $(p',2\zeta)$-biased over $\mathrm{GF}[2]$.  
    
    We now establish the claim of computation time. Let $\tilde{\mathbb{Z}} = \{2^{l} \, : \, l \in \mathbb{Z}_{\geq 0}\}$ and define $f:\mathbb{F} \times  \tilde{\mathbb{Z}} \to \mathbb{F}$ by $f(\lambda,p') := \sum_{i=0}^{p'-1} \lambda^{i}$. Observe that 
    \begin{equation}
        f(\lambda,p') = \sum_{i=0}^{p'-1} \lambda^{i} = (1+\lambda)(1+\lambda^{2} + \lambda^{4} + ... + \lambda^{p'-2}) = (1+\lambda)\sum_{i=0}^{p'/2 -1 } (\lambda^{2})^{i} = (1+\lambda)f(\lambda^{2},p'/2) \ .
    \end{equation}
    Writing $p' = 2^{l}$ for a positive integer $l$, we have $f(\lambda,2^{l}) = (1+\lambda)f(\lambda^{2},2^{l-1})$, and applying this procedure $l$ times, we arrive at
    \begin{equation}
        f(\lambda,p') = \prod_{j=0}^{l-1} (1+\lambda^{2^{j}})f(\lambda^{2^{l}},1) = \prod_{j=0}^{l-1} (1+\lambda^{2^{j}}) \ .
    \end{equation}
    For $\alpha \in A$, $\beta \in B$ and $\nu \in \mathbb{F}$, we can write
    \begin{equation}
        G(\beta,\nu)_{\alpha} = \nu \cdot \sum_{i=0}^{p'-1} (\alpha \beta)^{i} = \nu \cdot f(\alpha \beta,p') = \nu \cdot \prod_{j=0}^{l-1} (1+(\alpha \beta)^{2^{j}}) \ .
    \end{equation}
    The above element of $\mathbb{F}$, viewed as a string of $t$ bits, can be computed in $O(\log(p'))$ finite field operations over $\mathbb{F} = \mathrm{GF}[2^{t}]$. This completes the proof.
\end{proof}

\section{Proofs for the new construction of Raz's extractor}

\subsection{Proof of \cref{lem:newRaz1}} \label{app:newRaz1}
\noindent \textbf{Lemma 5}.
    \textit{Let $n_{1}$ and $n_{2}$ be positive integers, where $n_{1}$ is even and $n_{2} \leq n_{1}/2$. Define $N = (n_{1}/2)2^{n_{2}}$. Then for any positive integers $k_{1},k_{2},m,l,p$ and $\gamma > 0$ such that $m \leq n_{1}/2$, $l \leq n_{2} + \log(n_{1}/2), \ p \leq 2^{l}/m$, $p$ is even and any
    \begin{equation}
            \gamma \geq 2^{(n_{1}-k_{1})/p} \cdot \big[ (2\zeta)^{1/p} + p \cdot 2^{-k_{2}/2}\big],
        \end{equation}
    where $\zeta = 2^{l-n_{1}/2}$, we have the following:
    \begin{enumerate}[(i)]
        \item The generator of Construction~\ref{const:eps} viewed as a function $G:\{0,1\}^{n_{1}} \rightarrow \{0,1\}^{N}$ is a $(p',2\zeta)$-biased generator with $p' = 2^{l}$. 
        \item Consider the output of $G$ as $2^{n_{2}}$ blocks of size $n_{1}/2$ bits, and let $G(X)_{(i,y)}$ denote bit $i \in \{0,...,m-1\}$ of block $y \in \{0,1\}^{n_{2}}$. Then the function $\mathrm{Ext}:\{0,1\}^{n_{1}} \times \{0,1\}^{n_{2}} \rightarrow \{0,1\}^{m}$ defined by $\mathrm{Ext}(x,y)_{i} = G(x)_{(i,y)}$ is a $(n_1, k_1, n_2, k_2, m, \epsilon = 2^{m/2}\gamma)$ two-source extractor, and a strong (in either input) $(n_{1},k_{1},n_{2}',k_{2}',m, \gamma')$ two-source extractor, where
        \begin{equation}
            \begin{aligned}
                k_{1}' &= k_{1} + m/2 + 2 + \log (1/\gamma), \\
                k_{2}' &= k_{1} + m/2 + 2 + \log (1/\gamma), \\
                \gamma' &= \gamma \cdot 2^{m/2+1}.
            \end{aligned}
        \end{equation}
        \item Given $x \in \{0,1\}^{n_{1}}$ and $y \in \{0,1\}^{n_{2}}$, $\mathrm{Ext}(x,y)$ can be computed with computation time $O(n_{1}\log (n_{1})\log( p'))$.
    \end{enumerate} 
}
\begin{proof}
    \noindent \textbf{Part (i):} Choose the following parameters for \cref{lem:eff}:
    \begin{equation}
        t = n_{1}/2, \ n = 2^{n_{2}} \ .
    \end{equation}
    Since $\zeta = p' 2^{-n_{1}/2}$, $r = \log(p'/\zeta) = n_{1}/2$. We then have the finite field $\mathbb{F}$ is given by $\mathbb{F} = \mathrm{GF}[2^{t}] = \mathrm{GF}[2^{n_{1}/2}]$, and the two subsets are given by $A = \mathrm{GF}[n] = \mathrm{GF}[2^{n_{2}}]$ and $B = \mathrm{GF}[2^{r}] = \mathrm{GF}[2^{n_{1}/2}]$. The condition $2^{t} \geq \max\{n,2^{r}\}$ reads $2^{n_{1}/2} \geq \max \{ 2^{n_{2}}, 2^{n_{1}/2}\}$,
    which is satisfied by the constraint $n_{1}/2 \geq n_{2}$. We can therefore apply \cref{lem:eff} to obtain a generator for $n \cdot t = (n_{1}/2)2^{n_{2}}$ binary random variables $2p'2^{n_{1}/2}$-biased for linear tests of size $p'$, provided $p' \leq N$, which is satisfied since $l \leq n_{2} + \log (n_{1}/2)$. Note this sequence is generated using $t + r = n_{1}$ random bits.

    \vspace{0.2cm}

    \noindent \textbf{Part (ii):} We use the first $n_{1}/2$ bits of $X=x$ to select $\beta_{x} \in B = \mathrm{GF}[2^{n_{1}/2}]$, and the remaining $n_{1}/2$ bits to select $\nu_{x} \in \mathbb{F} = \mathrm{GF}[2^{n_{1}/2}]$. We can then define $2^{n_{2}}$ blocks, via arbitrary choice of $\alpha \in A \subset \mathrm{GF}[2^{n_{1}/2}]$, with $|A| = 2^{n_{2}}$, and associate the finite field elements $G(\beta_{x},\nu_{x})_{\alpha} \in \mathbb{F} = \mathrm{GF}[2^{n_{1}/2}]$, from \cref{eq:gen}. We denote its binary form 
    \begin{equation}
        G_{y}(x) = G(x)_{(y,0)},...,G(x)_{(y,n_{1}/2-1)} \in \{0,1\}^{n_{1}/2} \ ,
    \end{equation}
    where we exchanged the label $\alpha$ with a bitstring $y \in \{0,1\}^{n_{2}}$,\footnote{This assignment can be done using any bijection between the $2^{n_{2}}$ elements in $A$ and strings $y \in \{0,1\}^{n_{2}}$.} and exchanged the label $(\beta_{x},\gamma_{x})$ with $x$. By part $(i)$, the set $\{G(X)_{(y,i)}\}_{y,i}$ constitutes $(n_{1}/2)2^{n_{2}} \geq m \, 2^{n_{2}}$ binary random variables $2\zeta$-biased for linear tests of size $p'$, which can be constructed using $n_{1}$ random bits. The claim is then a corollary of \cref{lem:raz}. Note that only $m \, 2^{n_{2}}$ variables constructed from $n_{1}$ bits are required for \cref{lem:raz}, and the proof holds identically when we have access to $N > m \, 2^{n_{2}}$ bits since, given a string of $N$ bits $\zeta$-biased for linear tests of size $p'$, any sub-string of length $< N$ inherits the same property by definition. 

    \vspace{0.2cm}

    \noindent \textbf{Part (iii):} For a given $x \in \{0,1\}^{n_{1}}$ and $y \in \{0,1\}^{n_{2}}$, computation of the output $\mathrm{Ext}(x,y) = \mathrm{Ext}(x,y)_{0},...,\mathrm{Ext}(x,y)_{m-1}$ corresponds to computing the single block $G_{y}(x)$ and taking the first $m$ bits (since $m \leq n_{1}/2$). The claim then follows from the efficient properties of the generator $G$ in \cref{lem:eff}. 
\end{proof}

\subsection{Proof of \cref{thm:newRaz}}
\label{app:newRaz}
\noindent \textbf{Theorem 1.} \textit{Let $n_{1}, k_{1}, n_{2}, k_{2}, m$ be positive integers, $0 < \delta < 1/2$ and $0.25 < \lambda < (\delta k_2 / 16 - 1)$, such that $n_{2} \leq n_{1}/2$ and
    \begin{align}\label{constr3}
        k_1 &\geq \left(\frac{1}{2} + \delta\right)n_1 + 2\log(n_1) + 1\ , \\
        \label{constr4}
        k_2 &\geq \max\Big[3.2 \log \Big(\frac{8n_1}{k_2}\Big), 40 \Big]\ ,\\
       \label{constr5}
        m &\leq \frac{1}{\lambda} \Big( \frac{\delta k_2}{16} - 1 \Big)\ .
    \end{align}
    Then there exists an explicit $(n_{1},k_{1},n_{2},k_{2},m,\epsilon \leq 2^{(1 - 4\lambda)m/2 - 1})$ two-source extractor that can be computed in $O(n_1 \log (n_1)^2)$ time, and an explicit strong $(n_{1},k'_{1},n_{2},k'_{2},m,\epsilon' \leq 2^{(1 - 4\lambda)m/2})$ two-source extractor that can be computed in $O(n_1 \log (n_1)^2)$ time, with
    \begin{equation}
    \begin{aligned}
        k_1' &= k_1 + 3(m+1)\ , \\
        k_2' &= k_2 + 3(m+1)\ .
    \end{aligned}
    \end{equation} }

\begin{proof}
    Our proof follows the overall structure of Raz’s~\cite{Raz05} proof of Theorem 1, incorporating several improvements to achieve better parameters.
    We choose $N = (n_{1}/2) 2^{n_2}$, $p' = 2^{l}$ where $l = \floor*{\log \big(m(n_{1}-k_{1})\big)}$ and $\zeta = 2^{l-n_{1}/2}$. 
    Therefore, $\log(p')= O(\log(n_1))$ and $p' \leq m(n_{1}-k_{1})$. 
    Next, define $r := \log(p'/\zeta) = n_1/2$ and $t := n_1/2$. 
    Hence, by \cref{lem:eff}, we have that $G_0, \ldots, G_{N-1}$ binary random variables that are $2\zeta$-biased for linear tests of size $p'$ can be computed from $n_1$ random bits, with any constant multiple of $n_1/2$ being computable with time $O(n_1 \log (n_1)^2)$.
    For some even integer $p \leq p'$, we define 
    \begin{align}
        \log(\gamma_1) &= \frac{1}{p} \Big(n_{1}-k_{1} + \log(2\zeta) \Big) \ , \\
        \log(\gamma_2) &= (n_{1}-k_{1})/p + \log(p) - k_{2}/2 \ ,
        \label{eq:loggamma_2}
    \end{align}
    and
    \begin{align} 
        \gamma_1 + \gamma_2 = 2^{(n_{1}-k_{1})/p} \cdot \big[ (2\zeta)^{1/p} + p \cdot 2^{-k_{2}/2}\big]\ .
    \end{align}
    We now consider two cases. \newline
    \noindent \textbf{Case 1: $k_{2} < 4(n_{1}-k_{1})$.} Set $p$ to the smallest even integer larger than $8(n_{1}-k_{1})/k_{2}$.
    Then,
    \begin{align} \label{eq:pBounds}
        8(n_{1}-k_{1})/k_{2} \leq p \leq 8n_{1}/k_{2}\ .
    \end{align}
    Next, by inserting $\zeta = p' 2^{-n_{1}/2} = 2^{l - n_{1}/2}$ and $l = \floor*{\log m(n_{1}-k_{1})}$, we get
    \begin{equation}
    \begin{aligned}
        -\log(\gamma_{1}) &= \frac{-1}{p}\left(n_{1}-k_{1} + 1 + l - n_{1}/2 \right) \\
        &= \frac{1}{p}\left(k_{1} - n_{1}/2 - \floor*{\log  \big(m(n_{1}-k_{1})\big)}) -1 \right)\ .
    \end{aligned}
    \end{equation}
    To lower bound the above, we use the largest value of $p$ from \cref{eq:pBounds}, since \cref{constr3} implies that $k_1 - \frac{n_1}{2} - \floor*{\log \big(m(n_{1}-k_{1})\big)} - 1 \geq 0$. This follows from the fact that 
    \begin{equation}
    \begin{aligned}
        k_1 - \frac{n_1}{2} - \floor*{\log \big(m(n_{1}-k_{1})\big)} -1 &> k_1 - \frac{n_1}{2} - \log \big(\frac{k_2}{\lambda 32}(n_{1}-k_{1})\big) - 1\\ 
        &> k_1 - \frac{n_1}{2} - \log \big(\frac{k_2}{8}(n_{1}-k_{1})\big) -1 \\
        &> k_1 - \frac{n_1}{2} - 2\log(n_1) -1 \geq 0 \ ,
    \end{aligned}
    \end{equation} 
    where the first inequality uses \cref{constr5} followed by the bound $\delta < 1/2$, the second follows from the bound $\lambda > 1/4$, the third uses $k_{2}(n_{1} - k_{1}) < n_{1}^{2}$, and the last follows from \eqref{constr3} and the fact that $\delta > 0$.
    Therefore 
    \begin{align}
        -\log(\gamma_{1}) &= \frac{1}{p}\left(k_{1} - \frac{n_{1}}{2} - \floor*{\log \big(m(n_{1}-k_{1})\big)} -1\right) \\
        &\geq \frac{k_2}{8n_1}\left(k_{1} - \frac{n_{1}}{2} - \floor*{\log \big(m(n_{1}-k_{1})\big)} -1\right) \\
        &\geq \frac{k_2}{8n_1}\left(k_{1} - \frac{n_1}{2} - 2\log (n_1) -1\right) \\
        &\geq \frac{k_2}{8n_1}\delta n_1 \\
        &\geq 2 (\lambda m + 1) \ ,
    \end{align}
    where the penultimate inequality uses the bound on $k_1$ from \cref{constr3} and the final inequality uses the bound on $m$ from \cref{constr5}.
    Next, we bound $\gamma_2$ using the restrictions on $p$ in \cref{eq:pBounds},
    \begin{align}
        -\log(\gamma_{2}) &= \frac{k_{1} - n_1}{p} - \log(p) + \frac{k_{2}}{2} \\
        &\geq \frac{k_2(k_{1} - n_1)}{8(n_1 - k_1)} - \log\Big(\frac{8n_1}{k_2}\Big) + \frac{k_{2}}{2} \\
        &= \frac{3k_2}{8} - \log\Big(\frac{8n_1}{k_2}\Big)\ . 
    \end{align}
    Noting that $\log(8n_{1}/k_{2}) \leq 5k_{2}/16 $ by \eqref{constr4} and $\delta < 1/2$,
    \begin{equation}
        \frac{3k_2}{8} - \log\Big(\frac{8n_1}{k_2}\Big) \geq \frac{3k_2}{8} - \frac{5k_2}{16}
        > \frac{3k_2}{8} - \frac{3 - \delta}{8}k_{2} = \frac{k_{2}\delta}{8} \geq 2 (\lambda m + 1),
    \end{equation}
    where the final inequality comes from \cref{constr5}.
    Therefore, combining the bounds on $\gamma_{1}$ and $\gamma_{2}$, we get that
    \begin{align}
        \gamma_1 + \gamma_2 \leq 2^{-2\lambda m - 2} + 2^{-2\lambda m - 2} = 2^{-2\lambda m - 1}\ .
    \end{align} 
    \newline
    \noindent \textbf{Case 2: $k_{2} \geq 4(n_{1}-k_{1})$.} Set $p = 2$. 
    We find
    \begin{align}
        -\log(\gamma_{1}) &= \frac{1}{p}\left(k_{1} - \frac{n_{1}}{2} - \floor*{\log \big(m(n_{1}-k_{1})\big)} -1\right) \\
        &=  \frac{1}{2}\left(k_{1} - \frac{n_{1}}{2} - \floor*{\log \big(m(n_{1}-k_{1})\big)} -1\right) \\
        &\geq  \frac{k_2}{8n_1}\left(k_{1} - \frac{n_{1}}{2} - \floor*{\log \big(m(n_{1}-k_{1})\big)} -1\right) \\
        &\geq \frac{k_2}{8n_1}\delta n_1 \\
        &\geq 2 (\lambda m + 1) \ ,
    \end{align}
    where the first inequality follows from the constraint $k_{2} \leq n_{2} \leq n_{1}/2$, and the remainder follows the steps in Case 1.
    Now, we consider $\gamma_2$:
    \begin{equation}
    \begin{aligned}
       -\log(\gamma_{2}) &= \frac{k_{1} - n_1}{p} - \log(p) + \frac{k_{2}}{2} \\
        &= \frac{k_{1} - n_1}{2} - \log(2) + \frac{k_{2}}{2} \\
        &= \frac{k_{1} - n_1}{2} + \frac{k_{2}}{2} - 1 \\
        &\geq \frac{3k_{2}}{8} - 1 \\
        &\geq 2 (\lambda m + 1) \ ,
    \end{aligned} 
    \end{equation}
    where the penultimate inequality comes from the fact that the Case 2 condition implies $n_{1} - k_{1} \leq k_{2}/4$, so $(k_{1} - n_{1})/2 + k_{2}/2 -1 \geq - k_{2}/8 + k_{2}/2 - 1 = 3k_{2}/8 - 1$. The final inequality comes from the fact that $2(\lambda m + 1) \leq  \frac{\delta k_2}{8} \leq \frac{3k_{2}}{8} - 1$, where we used \eqref{constr5} followed by \eqref{constr4}.
    Therefore, combining the bounds on $\gamma_{1}$ and $\gamma_{2}$, we get that
    \begin{align}
        \gamma_1 + \gamma_2 \leq 2^{-2\lambda m - 2} + 2^{-2\lambda m - 2} = 2^{-2\lambda m - 1}\ .
    \end{align}
    By \cref{lem:newRaz1}, we now recover an $(n_{1},k_{1},n_{2},k_{2},m,\epsilon = 2^{m/2}\gamma \leq 2^{m/2}2^{-2\lambda m - 1} = 2^{(1 - 4\lambda)m/2 - 1})$ two-source extractor with computation time $O(n_{1} \log(n_{1})\log(p')) = O(n_{1}\log(n_{1})^{2})$,
    and a strong $(n_{1},k'_{1},n_{2},k'_{2},m,\epsilon' \leq 2^{(1 - 4\lambda)m/2})$ two-source extractor, where
    \begin{align}
        k_1' &= k_1 + 3(m+1) \ ,\\
        k_2' &= k_2 + 3(m+1)\ ,
    \end{align} with the same computation time, concluding the proof.
\end{proof}

\section{Quantum-proofing with the classical-quantum XOR lemma} \label{app:cq1}
Informally, the classical XOR lemma~\cite{GoldreichXOR} states that, given $m$ binary random variables $X = X_{0},...,X_{m-1}$, the statistical distance $\mathrm{SD}[X,U_{m}]$ is bounded by the maximum bias, $\mathrm{MB}[X]$, up to a factor of $2^{m/2}$, i.e. $\mathrm{SD}[X,U_{m}] \leq 2^{m/2 - 1} \mathrm{MB}[X]$. The maximum bias quantifies the uncertainty of sums of certain bit positions of $X$, $\mathrm{MB}[X] = 2\max_{\tau}\mathrm{SD}[X_{\tau},U_{1}]$, where $\tau$ is any non-empty sub-set of $\{0,...,m-1\}$ and $X_{\tau} = \bigoplus_{i \in \tau}X_{i}$. To extract many bits, Raz shows that the 1-bit extractor defined in \cite[Lemma 3.3]{Raz05} implies a bound on the maximum bias of the $m$-bit extractor output, $\mathrm{MB}[\mathrm{Ext}(X,Y)]$, from which the XOR lemma yields a bound on the uniformity of the full extractor output, $\mathrm{SD}[\mathrm{Ext(X,Y)},U_{m}]$, with a penalty factor $2^{m/2}$. One route to making the Raz extractor quantum-proof would be to take Raz's strong 1-bit extractor, and obtain a strong, quantum-proof 1-bit extractor via a general reduction such as the Markov~\cite{AF15} or bounded storage model~\cite{Konig_2008}. Then using a classical-quantum (cq) XOR lemma~\cite{Kasher}, an $m$-bit quantum-proof extractor could be obtained analogously to the original proof.    

This proof structure will, in general, give a different set of final parameters to those obtained in \cref{cor:qproofRaz}, where the Markov model was applied to the $m$-bit Raz extractor directly. Specifically, the additional extractor error incurred by applying the Markov model scales exponentially in the output size $m$ (cf. \cref{lem:qproofMM}), and only applying this to the 1-bit extractor may be less penalizing. However, for the cq-XOR lemma presented in~\cite{Kasher} we could not find an improvement; this is due to the fact that the cq version is not tight compared to its classical counter part, and contributes an additional factor of $2^{m/2}$ to the error (i.e., $\mathrm{SD}[X,U_{m}] \leq 2^{m - 1} \mathrm{MB}[X]$). This is enough to diminish any potential advantage from this alternative proof structure. On the other hand, if a cq-XOR lemma was proven with the same penalty as the classical version, an improvement in parameters would be possible. 

Formally, given a random string of $m$ bits $X_{0},...,X_{m-1}$, let $\tau \subseteq \{0,...,m-1\}$ be a non-empty subset of indices, and define the binary random variables $X_{\tau} = \bigoplus_{i \in \tau} X_{i}$. The maximum bias of $X$ is given by  
\begin{equation}
    \mathrm{MB}[X] := 2\cdot \max_{\tau \neq \varnothing} \mathrm{SD}[X_{\tau},U_{1}].
\end{equation}
Then the classical XOR lemma is the following result:
\begin{lemma}
    Let $X_{0},...,X_{m-1}$ be $m$ binary random variables. Then
    \begin{equation}
        \mathrm{SD}[X,U_{m}] \leq \frac{2^{m/2}}{2} \mathrm{MB}[X].
    \end{equation} \label{lem:cXOR}
\end{lemma}
\noindent For proof see~\cite{GoldreichXOR}. This relationship is established by relating $\mathrm{SD}$ and $\mathrm{MB}$ to the $l_{1}$ and $l_{\infty}$ norms, respectively, on the space of probability distributions in $\mathbb{R}^{2^{m}}$, and applying relevant norm inequalities. The quantum case is defined analogously: consider the cq-state
\begin{equation}
    \rho_{XE} = \sum_{x \in \{0,1\}^{m}} p_{X}(x)\ketbra{x}{x} \otimes \rho_{E}^{x}\ .
\end{equation}
Define, for $\mu \in \{0,1\}$, 
\begin{equation}
    \Pi_{\mu}^{\tau} := \sum_{\substack{y\in\{0,1\}^{m} \\ \mathrm{s.t.} \ \bigoplus_{i \in \tau}y_{i}=\mu}} \ketbra{y}{y}, \ K_{\mu}^{\tau} := \ket{\mu} \otimes  \Pi_{\mu}^{\tau}\ . 
\end{equation}
Note that $K_{0}^{\tau \dagger}K_{0}^{\tau} + K_{1}^{\tau \dagger}K_{1}^{\tau} = \Pi_{0}^{\tau} + \Pi_{1}^{\tau} = \mathds{1}_{X}$, hence $\{K_{\mu}^{\tau}\}_{\mu}$ form a set of Kraus operators for every non-empty $\tau$. Applying this channel to $\rho_{XE}$ yields
\begin{equation}
    \sum_{\mu =0}^{1} (K_{\mu}^{\tau} \otimes \mathds{1}_{E})\rho_{XE} (K_{\mu}^{\tau} \otimes \mathds{1}_{E})^{\dagger} = \sum_{\mu=0}^{1} \ketbra{\mu}{\mu} \otimes \Bigg( \sum_{\substack{x \in\{0,1\}^{m}\\ \mathrm{s.t.} \ \bigoplus_{i \in \tau}x_{i}=\mu}} p_{X}(x)\ketbra{x}{x} \otimes \rho_{E}^{x}\Bigg)\ .
\end{equation}
Taking the partial trace over $X$, and labeling the entire channel (including identity on $E$) $\Lambda_{\tau}$, we define
\begin{equation}
    \rho_{X_{\tau}E} := \Lambda_{\tau}[\rho_{XE}] = \sum_{\mu=0}^{1} \ketbra{\mu}{\mu} \otimes \Bigg( \sum_{\substack{x \in\{0,1\}^{m}\\ \mathrm{s.t.} \ \bigoplus_{i \in \tau}x_{i}=\mu}} p_{X}(x)\rho_{E}^{x}\Bigg)\ . 
\end{equation}

This allows us to define the maximum bias of $\rho_{XE}$ with respect to $E$ (slightly neglecting notation by using the same label as in the classical case), 
\begin{equation}
    \mathrm{MB}[\rho_{XE}]:= \max_{\tau \neq \varnothing}  \| \rho_{X_{\tau}E} - \omega_{2} \otimes \rho_{E}\|_{1}\ ,
\end{equation}
where the maximum is taken over all non-empty subsets $\tau$. The aim is to bound $\mathrm{TD}[\rho_{XE},\omega_{m} \otimes \rho_{E}]$ by $\mathrm{MB}[\rho_{XE}]$; we interpret such a bound as a classical-quantum version of the XOR lemma. Notably, a cq-XOR lemma of this type was proven by Kasher and Kempe~\cite{KK}:
\begin{lemma}[\cite{KK}, Lemma 10]
    Let $\rho_{XE}$ be an arbitrary cq-state, and $d = \mathrm{dim}[\mathcal{H}_{E}]$. Then
    \begin{equation}
        \mathrm{TD}[\rho_{XE},\omega_{m} \otimes \rho_{E}] \leq 2^{\min\{m,d\}/2}\sqrt{\sum_{\tau \neq \varnothing} \Big( \mathrm{TD}[\rho_{X_{\tau}E} , \omega_{2} \otimes \rho_{E}] \Big)^{2} }. \label{eq:oldcqXOR}
    \end{equation} \label{lem:oldCQXOR}
\end{lemma}
\noindent \cref{lem:oldCQXOR} can be used to quantum-proof Raz's extractor in the following way. Recall Raz's strong, 1-bit extractor, obtained by setting $m =1$ in \cref{lem:raz}.
\begin{lemma}[\cite{Raz05}]
    Let $N = 2^{n_{2}}$. Let $G_{0},...,G_{N-1}$ be 0-1 random variables $\zeta$-biased for linear tests of size $p'$ that can be constructed using $n_{1}$ random bits. Define $\mathrm{Ext}:\{0,1\}^{n_{1}}\times \{0,1\}^{n_{2}} \to \{0,1\}$ by $\mathrm{Ext}(x,y) = G(x)_{y}$. Then, for any even integer $p \leq p'$ and any $k_{1},k_{2}$, the function $\mathrm{Ext}$ is an $(n_{1},k_{1}',n_{2},k_{2}',1,\gamma \, 2^{3/2})$ strong (in either source) two-source extractor for any $\gamma \geq 2^{(n_{1}-k_{1})/p} \cdot \big[ \zeta^{1/p} + p \cdot 2^{-k_{2}/2}\big]$ and $k_{1}' = k_{1} + 5/2 + \log(1/\gamma)$, $k_{2}' = k_{2} + 5/2 + \log(1/\gamma)$.  
\end{lemma}
\noindent Applying \cref{lem:qproofMM}, we obtain a strong 1-bit two-source extractor that is quantum-proof in the Markov model,
\begin{corollary}
    Let $N = 2^{n_{2}}$. Let $G_{0},...,G_{N-1}$ be 0-1 random variables $\zeta$-biased for linear tests of size $p'$ that can be constructed using $n_{1}$ random bits. Define $\mathrm{Ext}:\{0,1\}^{n_{1}}\times \{0,1\}^{n_{2}} \to \{0,1\}$ by $\mathrm{Ext}(x,y) = G(x)_{y}$. Then, for any even integer $p \leq p'$ and any $k_{1},k_{2}$, the function $\mathrm{Ext}$ is a quantum-proof $(n_{1},k_{1},n_{2}',k_{2}',1,\sqrt{3\sqrt{2}\gamma})$ two-source extractor in the Markov model, for any $\gamma \geq 2^{(n_{1}-k_{1})/p} \cdot \big[ \zeta^{1/p} + p \cdot 2^{-k_{2}/2}\big]$ and $k_{1}' = k_{1} + 1 + 2\log(1/\gamma)$, $k_{2}' = k_{2} + 1 + 2\log(1/\gamma)$. \label{cor:1bitQ}
\end{corollary}

\noindent Combining with \cref{lem:oldCQXOR}, we arrive at the following:
\begin{lemma}
    Let $N = m \cdot 2^{n_{2}}$. Let $G_{0},...,G_{N-1}$ be 0-1 random variables $\zeta$-biased for linear tests of size $p'$ that can be constructed using $n_{1}$ random bits. Define $\mathrm{Ext}:\{0,1\}^{n_{1}}\times \{0,1\}^{n_{2}} \to \{0,1\}^{m}$ by $\mathrm{Ext}(x,y)_{i} = G(x)_{(i,y)}$. Then, for any even integer $p \leq p'/m$ and any $k_{1},k_{2}$, the function $\mathrm{Ext}$ is a strong (in either input) quantum-proof $(n_{1},k_{1},n_{2}',k_{2}',m,2^{m} \, \sqrt{3\sqrt{2}\gamma})$ two-source extractor in the Markov model, with $\gamma \geq 2^{(n_{1}-k_{1})/p} \cdot \big[ \zeta^{1/p} + p \cdot 2^{-k_{2}/2}\big]$ and $k_{1}' = k_{1} + 1 + 2\log(1/\gamma)$, $k_{2}' = k_{2} + 1 + 2\log(1/\gamma)$. \label{lem:mbitQ} 
\end{lemma}
\begin{proof}
    First, we reproduce the proof in \cite[Lemma 3.3]{Raz05}. For every non-empty $\tau \in \{0,...,m-1\}$, define $\mathrm{Ext}_{\tau} : \{0,1\}^{n_{1}} \times \{0,1\}^{n_{2}} \to \{0,1\}$, by 
    \begin{equation}
        \mathrm{Ext}_{\tau}(x,y) = \bigoplus_{i \in \tau} \mathrm{Ext}_{i}(x,y) = \bigoplus_{i \in \tau} Z_{(i,y)}(x)\ .
    \end{equation}
For a fixed $\tau$, note that the set of variables $\big \{ Z_{y|\tau} := \bigoplus_{i \in \tau} Z_{(i,y)} \ : \ y \in \{0,1\}^{n_{2}}\}$ is $\zeta$-biased for linear tests of size $p'/m$. This follows from the fact that, for any non-empty $\mathcal{Y} \subset \{0,1\}^{n_{2}}$ satisfying $|\mathcal{Y}| \leq p'/m$, define  
\begin{equation}
    Z_{\mathcal{Y}} := \bigoplus_{y \in \mathcal{Y}} Z_{y|\tau} = \bigoplus_{y \in \mathcal{Y}}\bigoplus_{i \in \tau} Z_{(i,y)}\ .
\end{equation}
Note that $|\mathcal{Y}|\cdot | \tau | = p' |\tau|/m \leq p'$, implying $Z_{\mathcal{Y}}$ must be $\zeta$-close to uniform, since $\{Z_{(i,y)}\}_{i,y}$ are $\zeta$-biased for linear tests of size $p'$. Therefore, for every non-empty $\tau$, the variables $\{ \mathrm{Ext}_{\tau}(x,y)\}_{y \in \{0,1\}^{n_{2}}} = \{ Z_{y|\tau}(x) \}_{y \in \{0,1\}^{n_{2}}}$ are $\zeta$-biased for linear tests of size $p'/m$. By \cref{cor:1bitQ} the function $\mathrm{Ext}_{\tau}(x,y) = Z_{y|\tau}(x)$ is an $(n_{1},k_{1},n_{2}',k_{2}',1,\epsilon'=\sqrt{3\gamma/2})$ strong (in either input) 1-bit extractor, quantum-proof in the Markov model from \cref{cor:1bitQ}. Choosing the extractor to be strong in the first source, this implies
    \begin{equation}
        \mathrm{TD}\big[\rho_{\mathrm{Ext}_{\tau}(X,Y)XE}, \omega_{2} \otimes \rho_{XE} \big] \leq \epsilon' \label{eq:exp}
    \end{equation} 
    for all non-empty $\tau$. Notice that
    \begin{equation}
        \rho_{\mathrm{Ext}_{\tau}(X,Y)XE} = \Lambda_{\tau}[\rho_{\mathrm{Ext}(X,Y)XE}]\ ,
    \end{equation}
    and \cref{eq:exp} implies $\mathrm{MB}[\rho_{\mathrm{Ext}(X,Y)XE}] \leq 2 \epsilon'$. We can directly apply \cref{lem:oldCQXOR} to obtain
    \begin{equation}
        \mathrm{TD}[\rho_{\mathrm{Ext}(X,Y)XE},\omega_{m}\otimes \rho_{XE}] \leq \epsilon' \cdot 2^{m}\ .
    \end{equation}
\end{proof}
Notice that the error increases by a factor of $2^{m}$, compared to $2^{m/2}$ in the classical case. Alternatively, we could apply the Markov model directly to the original $m$-bit extractor, resulting in the parameters of \cref{cor:mbitQ2}, restated below:

\vspace{0.2cm}

\noindent \textbf{Corollary 2.} \textit{Let $N = m \cdot 2^{n_{2}}$. Let $G_{0},...,G_{N-1}$ be 0-1 random variables $\zeta$-biased for linear tests of size $p'$ that can be constructed using $n_{1}$ random bits. Define $\mathrm{Ext}:\{0,1\}^{n_{1}}\times \{0,1\}^{n_{2}} \to \{0,1\}^{m}$ by $\mathrm{Ext}(x,y)_{i} = G(x)_{(i,y)}$. Then, for any even integer $p \leq p'/m$ and any $k_{1},k_{2}$, the function $\mathrm{Ext}$ is a strong (in either input) $(n_{1},k_{1}',n_{2},k_{2}',m,2^{3m/4} \, \sqrt{3\gamma/2})$ two-source extractor quantum-proof in the Markov model, with $\gamma \geq 2^{(n_{1}-k_{1})/p} \cdot \big[ \zeta^{1/p} + p \cdot 2^{-k_{2}/2}\big]$ and $k_{1}' = k_{1} + 1 + 2\log(1/\gamma)$, $k_{2}' = k_{2} + 1 + 2\log(1/\gamma)$.}

\noindent Comparing the parameters, we have 
    \begin{align}
        \mathrm{\cref{lem:mbitQ} \ (CQ-XOR \ lemma):}& \ \ \ \epsilon = 2^{m} \, \sqrt{3\sqrt{2}\gamma}, \ \ \ k_{i}' = k_{i} + 1 + 2\log(1/\gamma), \label{eq:q1}\\
        \mathrm{\cref{cor:mbitQ2} \ (Markov \ model \ directly):}& \ \ \ \epsilon = 2^{3m/4} \, \sqrt{3\gamma/2}, \ \  k_{i}' = k_{i} + 1 + 2\log(1/\gamma).\label{eq:q2}
    \end{align}
On the other hand, a tight CQ-XOR lemma would result in 
\begin{equation}
    \mathrm{(Tight \ CQ-XOR \ lemma):} \ \ \ \epsilon = 2^{m/2} \, \sqrt{3\sqrt{2}\gamma}, \ \ \ k_{i}' = k_{i} + 1 + 2\log(1/\gamma), \label{eq:q3}
\end{equation}
which would give an improvement by decreasing the error exponent from $3m/4$ to $m/2$.

\end{document}